\documentclass[]{tMPH2e}

\usepackage{epsfig,epic,eepic,color,rotate,graphics,fancybox}
\usepackage{graphicx}
\usepackage{epstopdf}
\usepackage{color}
\begin{document}



\title{{\itshape The Chiral Dipolar Hard Sphere Model.}}
\author{Martial Mazars $^{\ast}$\thanks{$^\ast$Corresponding author. Email: Martial.Mazars@th.u-psud.fr
\vspace{6pt}}\\\vspace{6pt}{\em{Laboratoire de Physique Th\'eorique (UMR 8627), Universit\'e de Paris Sud XI, B\^atiment 210, 91405 Orsay Cedex, FRANCE}}.
\\\vspace{6pt}
\received{} }

\maketitle

\begin{abstract}
A simple molecular model of chiral molecules is presented in this paper : the chiral dipolar hard sphere model. The discriminatory interaction between enantiomers is represented by electrostatic (or magnetic) dipoles-dipoles interactions : short ranged steric repulsion are represented by hard sphere potential and, in each molecule, two point dipoles are located inside the sphere. The model is described in detail and some of its elementary properties are given ; in particular, it is shown that the that the knowledge of only three multipole spherical components (namely : $Q_{10}$, $Q_{21}$ and $Q_{22}$) allows to compute all multipole spherical components of the model. Despite, the simplicity of the model, it is shown also that the energy landscape of the interaction between two enantiomers is quite rich, this renders systems of chiral dipolar hard sphere very interesting and complicated to study. Few preliminary Monte Carlo simulation results are also reported in the paper. Last, but not least, this paper is dedicated to Jean-Jacques Weis. 
\end{abstract}
\section{\label{sec:level1} Introduction}
Chirality is used to describe a lack of symmetry in compounds : it refers to objects, molecules or structures that are not superposable on their mirror image. The separation of the two optical active forms of the tartaric acid, based on the observation of the symmetry of crystals by Pasteur in 1847, has been a major step in the understanding of optical properties of crystals and liquids \cite{Pasteur:1847}. Later, in 1884, Kelvin introduced the term chirality, but it has been extensively and systematically used only eighty years later with the introduction of the Cahn-Ingold-Prelog priority rules \cite{Cahn:56}. Since Arago, Biot and Pasteur, the ideas and concepts on optical activity of substances and on chirality have had and still have a lot of applications in chemistry, biology and physics.\\
In chemistry and biochemistry, chirality is an important tool in stereochemistry and for the determination of the functions of molecules. The interactions between biological molecules are frequently a consequence of the chiral interactions between them. For instance, as first noted by Pasteur, the two forms of asparagine ({\it dextro} and {\it levo}) tasted differently ; it was found later that this is a frequent property the two optical active forms of amino acids, because of the chiral structure of the taste bud receptor site (several others interesting examples may be found in \cite{Craig:76}). The helical structures of DNA, RNA, proteins, etc. are also important to permit biological recognition and interactions between biological structures \cite{Kornyshev:07,Kuhn:08}.\\
In condensed matter physics many properties depend on the chirality of constituants or structures. More specifically, in liquid crystal physics, many phases as cholesteric, some smectics phases ($S^*_A$, $S^*_C$, etc.) and blue phases possess an optical rotatory power \cite{Meyer:75,vanderMeer:76,Goossens:89,Chandrasekhar:92,deGennes:93,Kamien:01,Emelyanenko:06} ; in most cases, a helical arrangement of the molecules is responsible for this property. Obviously, these properties are of a particular interest for technological applications. It is not necessary for  the molecules to be chiral to observe chiral liquid crystal phases \cite{deGennes:93,Xu:01,Yan:08} ; however, some particular phases, such as the ferroelectric smectic-$C^*$ phase, may be observed only when some chiral molecules are present in the sample (see section 5.10 of ref.\cite{Chandrasekhar:92} and refs.\cite{deJeu:03,Osipov:96,Cepic:06}).\\
To describe the optical activity of liquids and solids, and the discriminatory interaction between chiral molecules several molecular models have been built. Among the most known are Kuhn models that are based on coupled-oscillators arranged on an asymmetric tetrahedron \cite{Kuhn:30,Caldwell:71}, the optical activity tensor or gyration polarizability may be computed in these models \cite{Osipov:95,Wang:05,Wang:08}. Models based on chirality of helix have also been used \cite{Maki:96,Wang:05,Trost:07}. These helical models are important for the study of the properties of carbon nanotubes \cite{Barros:06,Anantram:06,Charlier:07} and biological macromolecules \cite{Kornyshev:07,Lee:04,Grason:07}.\\
For the study of chirality in fluids and solids, the discriminatory interactions between enantiomers are important too \cite{Craig:76}. The difference in the interaction between enantiomeric molecules may have several origins, it may stems from steric, electrostatic, magnetic or dispersion interactions \cite{Craig:76,Barron:87,Jenkins:94}. These {\it chirodiastaltic} interactions \cite{note1} are responsible for some differences in thermodynamical and structural properties between enantiopure systems (homochiral systems - systems that are composed with exclusively one enantiomer) and racemic systems (systems composed in equal proportions of the two enantiomers). For instance, the melting point of crystals of the enantiomeric pure $(+)$-tartaric acid is $170^o$C, while, for the racemic crystals, the melting temperature is $204-6^o$C. Some differences are also reported for boiling points, but since the chirodiastaltic interactions are quite small in general, these difference in boiling points are small and do not occur systematically. In liquid crystals, the phase diagrams depend also on the fractional concentrations of enantiomers \cite{Yang:87}.\\
Several simple molecular models have been proposed and used to study the thermodynamical and structural properties of fluids and solids of chiral molecules in computer simulations and integral equation theories. In these simple molecular models, the chirodiastaltic interactions are represented either via steric interactions \cite{Evans:92,Ferrarini:96,Berardi:03,Cao:05,Peon:06} or electrostatic interactions \cite{Paci:01,Paci:03,Paci:04,Huh:04} ; a chiral Gay-Berne model has also been built \cite{Memmer:01}. In the present work, we describe another simple model of chiral molecules, the chiral dipolar hard sphere model. In this model, the chirodiastaltic interactions is represented by electrostatic (or magnetic) dipoles-dipoles interactions, it consists of hard sphere with two point dipoles located inside the molecule. As shown by Craig and coworkers, electrostatic interactions between enantiomers may be discriminatory at most if the dipole-quadrupole interaction between two enantiomers differs \cite{Craig:76,Craig:74,Craig:75} ; this implies that if the quadrupole moment of the two form of enantiomers differs then the interaction may be discriminatory. Therefore, a chiral model based on electrostatic interactions must have at least four point electric charges arranged on an asymmetric tetrahedron geometry, as the models in refs.\cite{Paci:01,Paci:03,Paci:04,Huh:04}, two non-colinear point dipoles \cite{note2} or one quadrupole with no planar symmetry \cite{Craig:74}.\\ 
The present paper is organised as follow. In section 2, we describe the model of chiral dipolar hard sphere in detail and give some of its elementary properties. More precisely, in section 2.1, we define the model, the nomenclature, its pertinent parameters and we compute all its multipole moments ; it is shown that the knowledge of only three multipole spherical components (namely : $Q_{10}$, $Q_{21}$ and $Q_{22}$) allows to compute all multipole spherical components of the model. The knowledge of all multipole moments is important in the definition of the model and in the determination of its electrical properties and also, in view of the determination of the optical activity of the model, for the computation of multipole-multipole polarizabilities \cite{Craig:76,Caldwell:71,Craig:74,Craig:75,Luzanov:01,Bosello:03,Salam:06,Choi:07}. In section 2.2, we compute interaction energies between two enantiomers in some particular configurations ; it is shown that, despite the simplicity of the molecular model, the energy landscape of the interaction between two enantiomers is quite complicated. In section 3, we report some preliminary results of Monte Carlo simulations performed in the canonical $(NVT)$ and isobaric $(NPT)$ ensembles ; in these computations, all dipole-dipole interactions between molecules are explicitly taken into account by using the Ewald method for dipolar interactions \cite{Weis:05c}. The paper ends with a discussion on the perspectives that this model offers to the study of chiral fluids and solids. For completeness, in appendix A, we give general formulas for the computation of multipole moments for any dipole distribution.\\ 
Last, but not least, this paper is dedicaded to my colleague Jean-Jacques Weis.
\section{\label{sec:level2} Elementary properties of chiral dipolar hard sphere model.}
In this section, we describe the chiral dipolar hard sphere model and we give some of its elementary properties. The subsection 2.1 is devoted to the definition of the model and to the computation of the multipole moments due to the two permanent dipoles moments ; all multipole moments are computed with the center of the sphere taken as origin.\\
In subsection 2.2, we compute interaction energies between two enantiomers for few particular configurations ; in this subsection, a particular attention is pay to the {\it chirodiastaltic} interactions  for colinear aligned total dipole moments and antiparallel dipole moment configurations. The discriminatory interaction energies are often computed by using the multipole expansion \cite{Craig:76, Craig:74, Paci:01, Paci:04,Luzanov:01,Salam:06,Trost:07,Choi:07,Wang:08} ; thus, to facilitate the comparison with these works, the discriminatory interaction energies between chiral dipolar hard sphere have also been computed analytically by using the multipole expansion when the distance between both enantiomers is large in comparison to their diameter. However, if both enantiomers are at contact or if the distance between them is too small to use with accuracy the multipole expansion, then the interaction energies must be computed by summing all dipole-dipole interactions ; it is done so in subsection 2.2 for all configurations at contact and in section in all Monte-Carlo computations done with the Ewald method.  
\subsection{Definition of the model and its multipole moments.}
\begin{figure}
\centerline{\includegraphics[width=2.5in]{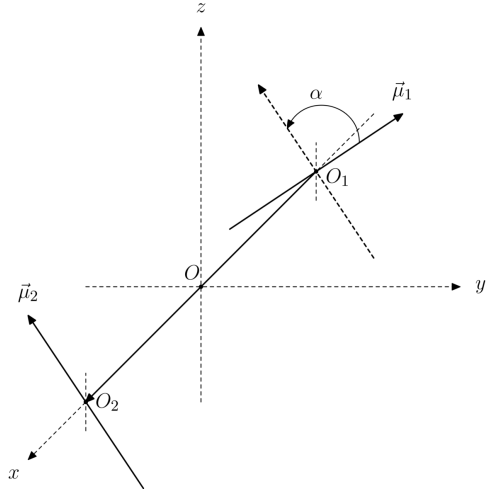}}
\caption{{\bf } Representation the molecule-fixed local frame for the Sinister bidipolar hard sphere enantiomer (S$_\alpha$-HS). The frame is defined from vectors $\hat{\bm{u}}$, $\bm{\mu}_1$ and  $\bm{\mu}_2$ via Eqs.(\ref{Dipole1}-\ref{frame}) ; $\mid \bm{O}_1\bm{O}_2\mid = 2L<\sigma$. The thick dashed vector is the projection of $\bm{\mu}_2$ in the plane $(O_1,\hat{\bm{e}}_z,\hat{\bm{\mu}}_1)$.}
\label{Fig1}
\end{figure}

The Chiral Dipolar Hard Sphere model consist of hard spheres of diameter $\sigma$ with two point dipoles $\bm{\mu}_1$ and  $\bm{\mu}_2$ located at points $O_1$ and $O_2$ inside the sphere. We define the molecular axis $\hat{\bm{u}}$ of a chiral dipolar hard sphere as $\bm{O}_1\bm{O}_2=2L \hat{\bm{u}}$ with $2L<\sigma $. The dipoles $\bm{\mu}_1$ and  $\bm{\mu}_2$ are chosen such that $\bm{\mu}_a.\hat{\bm{u}}=0$ and we set
\begin{equation}
\left\{ \begin{array}{ll}
\displaystyle \bm{\mu}_1&\displaystyle = \mu \mbox{ }\hat{\bm{\mu}}_1\\
&\\
\displaystyle \bm{\mu}_2&\displaystyle = \lambda\mu \left(\cos\alpha \mbox{ } \hat{\bm{\mu}}_1+ \sin\alpha \mbox{ }( \hat{\bm{\mu}}_1\times\hat{\bm{u}})\right)
\end{array}
\right.
\label{Dipole1}
\end{equation}
where the parameters $\alpha\in[-\pi,\pi]$ and $0\leq\lambda\leq1$ define the second dipole $\bm{\mu}_2$ from the dipole $\bm{\mu}_1$ and the axis $\hat{\bm{u}}$. We name {\it Rectus (R-) bidipolar hard sphere\/} (R$_\alpha$-HS) for $0<\alpha<\pi$ and {\it Sinister (S-) bidipolar hard sphere\/} (S$_\alpha$-HS) for $-\pi < \alpha <0$ (see figure \ref{Fig1} ). In this nomenclature, it is clear that R$_{(-\alpha)}$-HS $\equiv$ S$_\alpha$-HS ; molecules R$_\alpha$-HS and S$_\alpha$-HS are enantiomers.\\
Chiral parameters or chiral indices have been introduced to give a quantitative measure of the chirality of a molecule \cite{Osipov:95, Harris:99}. In particular, A.B. Harris and co-workers \cite{Harris:99} have defined chiral parameters as pseudoscalars constructed from the structure of molecules described via the representation theory of the three-dimensional rotation group $O(3)$ ;  they have also related chiral parameters to the cholesteric pitch of generic liquid crystal models. For the chiral dipolar hard sphere model, we may defined a pseudoscalar from the three vectors $\hat{\bm{u}}$, $\bm{\mu}_1$ and  $\bm{\mu}_2$ as 
\begin{equation}
\displaystyle\chi=L\hat{\bm{u}}.(\bm{\mu}_2\times\bm{\mu}_1)=L\lambda\mu^2\sin\alpha
\label{ChiralP}
\end{equation}
and take $\chi$ as the chiral parameter of this model ; we have $\chi(\mbox{S}_\alpha-HS)=-\chi(\mbox{R}_\alpha-HS)$. If $\chi=0$, then both enantiomers are equivalent and the bidipolar hard sphere model is achiral. This occurs if one of the following conditions is verified : (a) $L=0$, then both point dipoles are located at the center of the sphere and the model reduces to the dipolar hard sphere model ; (b) $\lambda=0$, in this case, there is only one point dipole in the molecule and (c) $\sin\alpha=0$ then the dipoles and the molecular axis are all in the same plane which is a plane of symmetry for the molecule.\\
A physical interpretation to $\mid \chi^*\mid =1$ is less obvious ; to obtain $\mid \chi^*\mid =1$ one needs to have $\lambda=1$, thus a symmetry is restored and then the chirality is perhaps less marked for these particular cases with $\lambda=1$ (for instance, see below subsection 2.2 and Fig.7). A trivial interpretation of values $\mid \chi^*\mid =1$ would be to say that for these value a maximum discrimination is achieved ; however, this is perhaps not fully correct.\\ 
We define the molecule-fixed local frame as 
\begin{equation}
\left\{ \begin{array}{ll}
\displaystyle \hat{\bm{e}}_z&\displaystyle = \frac{\bm{\mu}_1+\bm{\mu}_2}{\mid \bm{\mu}_1+\bm{\mu}_2\mid} = \frac{\bm{P}}{\mid\bm{P}\mid}\\
&\\
\displaystyle \hat{\bm{e}}_x&\displaystyle = \hat{\bm{u}}\\
&\\
\displaystyle \hat{\bm{e}}_y&\displaystyle =  \hat{\bm{e}}_z\times\hat{\bm{u}}
\end{array}
\right.
\label{frame}
\end{equation}
where $\bm{P}$ is the total dipole of the molecule. In this molecule-fixed frame, the projections of the dipoles onto the frame axis are given by 
\begin{figure}
\centerline{\includegraphics[width=5in]{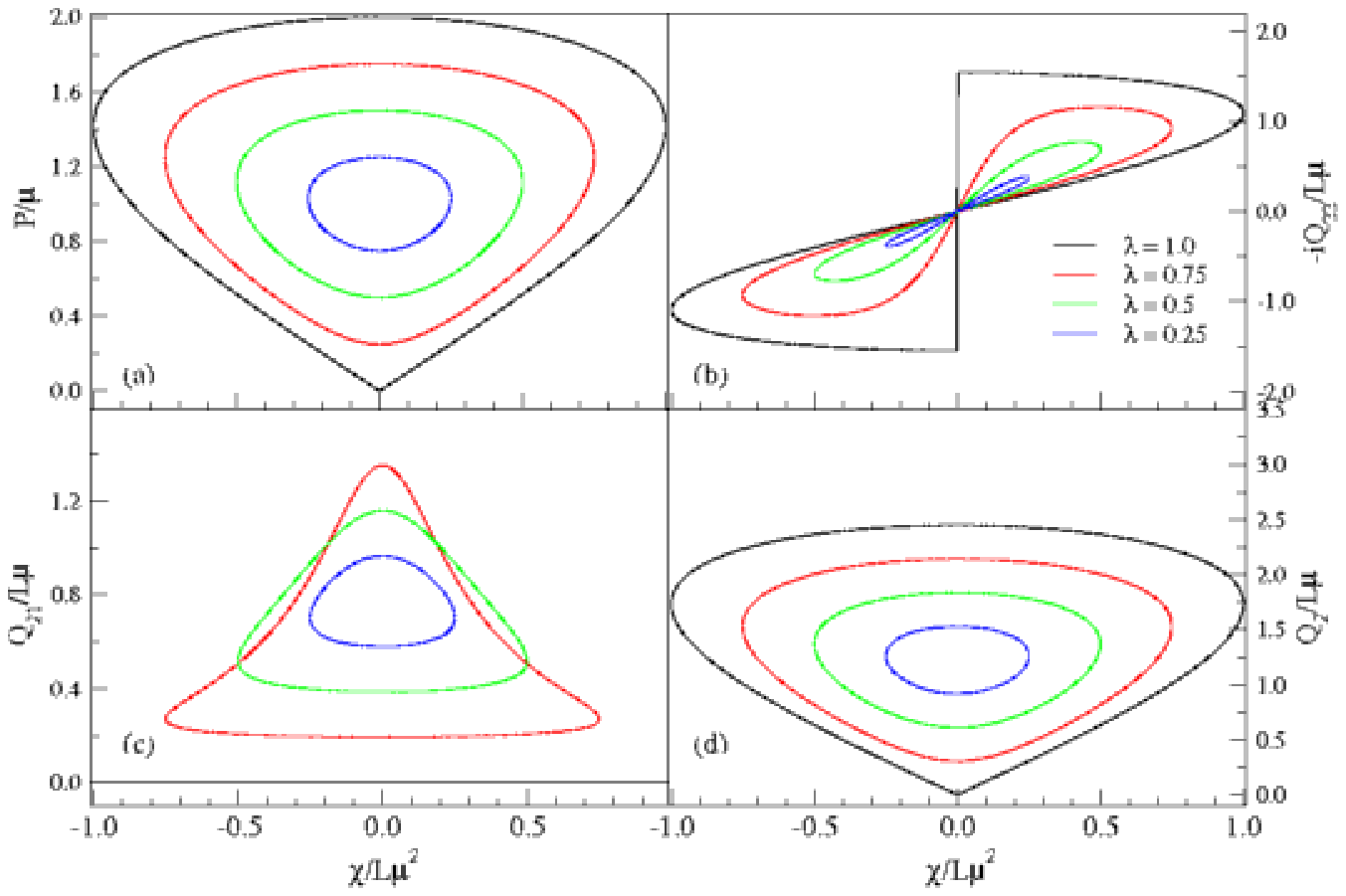}}
\caption{{\bf } Representation of the spherical components of the dipole and quadrupole moments as function of the chiral parameter $\chi$. The spherical components, represented as function of $\chi$, are multivalued because of the proprety $\chi(\alpha)=\chi(\pi-\alpha)$ (cf.Eq.(\ref{ChiralP})). Enantiomers R$_\alpha$-HS and S$_\alpha$-HS have equal spherical components when an even symmetry occurs, that is to say for $P$ and $Q_{21}$. One may note that for $\lambda=1$ one has $Q_{21}=0$.}
\label{Fig2}
\vspace{0.2in}
\centerline{\includegraphics[width=5in]{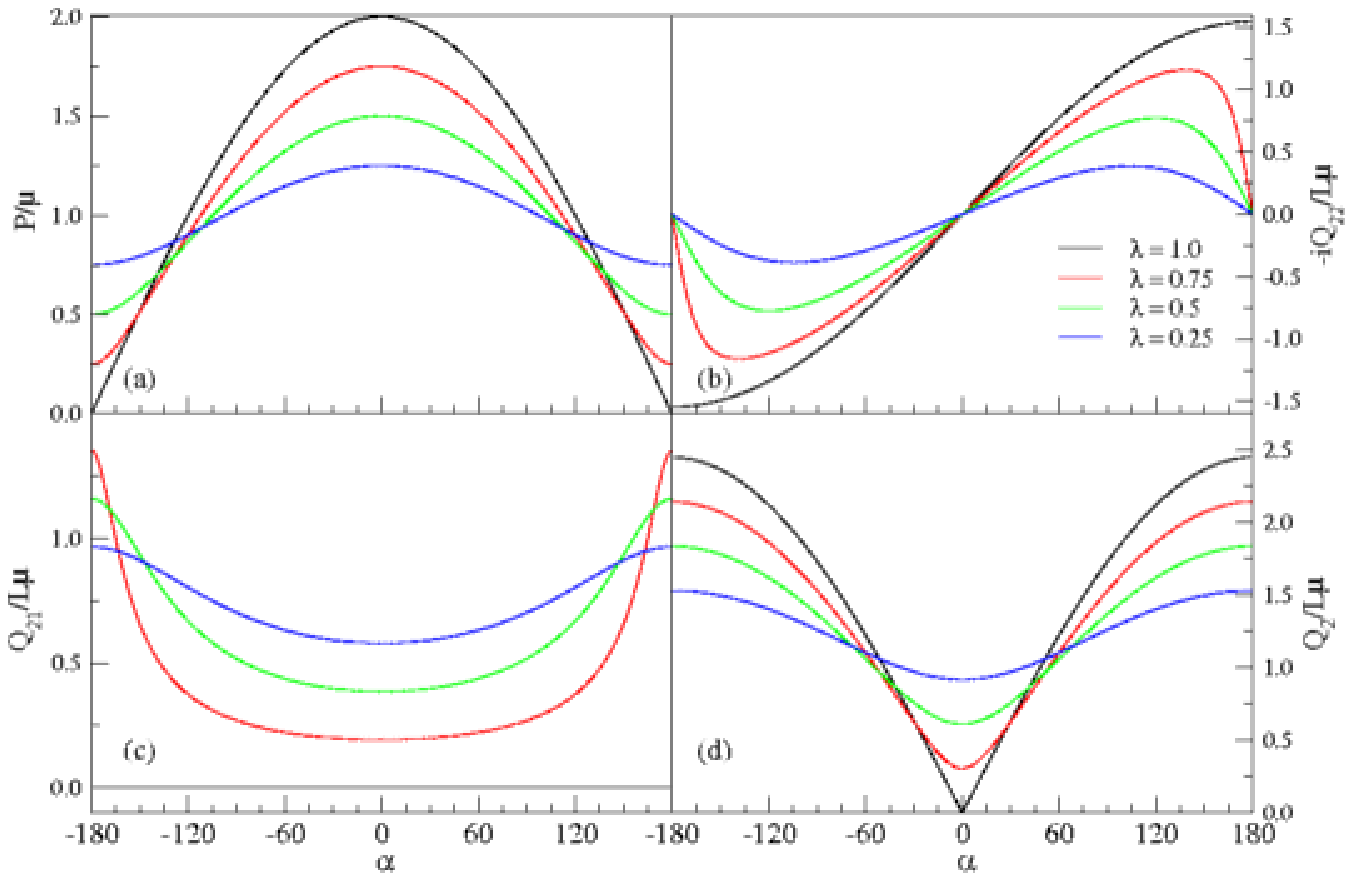}}
\caption{{\bf } Representation of the spherical components of the dipole and quadrupole as function of $\alpha$. As shown below, the knowledge of $P$, $Q_{21}$ and $Q_{22}$ as function of the parameters of the model determine fully all components of the spherical multipole tensor of an arbitrary order $l$ (see below).}
\label{Fig3}
\end{figure}
\begin{equation}
\left\{ \begin{array}{ll}
\displaystyle\bm{\mu}_1&\displaystyle = \mu (1+2\lambda\cos\alpha +\lambda^2)^{-1/2}\left[ -\lambda\sin\alpha \mbox{ } \hat{\bm{e}}_y+ (1+\lambda\cos\alpha)\mbox{ } \hat{\bm{e}}_z\right]\\
&\\
\displaystyle \bm{\mu}_2&\displaystyle = \mu (1+2\lambda\cos\alpha +\lambda^2)^{-1/2}\left[ \lambda\sin\alpha \mbox{ } \hat{\bm{e}}_y+ \lambda(\cos\alpha +\lambda) \mbox{ } \hat{\bm{e}}_z\right]\\
&\\
\displaystyle \bm{P}&\displaystyle = \mu (1+2\lambda\cos\alpha +\lambda^2)^{1/2} \mbox{ } \hat{\bm{e}}_z
\end{array}
\right.
\label{dipole}
\end{equation}
It is worthwhile to note that the total dipole of enantiomers R$_\alpha$-HS and S$_\alpha$-HS in the molecule-fixed frame are the same, we have $\bm{P}(\mbox{R$_\alpha$-HS})\equiv\bm{P}(\mbox{S$_\alpha$-HS})$. One should note also that $\bm{P}=0$ can be obtained only if $\lambda=1$ and $\alpha=\pi$, then $\chi=0$ and the model is achiral ; thus, all chiral bi-dipolar hard sphere have a non-zero net total dipole.\\
The dipole distribution of the chiral hard sphere may be written as
\begin{equation}
\bm{\mu}(\bm{r})= \bm{\mu}_1\delta(\bm{r}+L\hat{\bm{u}})+\bm{\mu}_2\delta(\bm{r}-L\hat{\bm{u}})
\label{disdip}
\end{equation}
and from multipole expansion for any dipolar distribution (see appendix A), the cartesian components of the quadrupole moment are given by
\begin{equation}
 \displaystyle [q]=\frac{3}{2}\frac{L\mu}{\sqrt{1+2\lambda\cos\alpha +\lambda^2}}\left[ \begin{array}{ccc}
 0 & 2\lambda\sin\alpha & (\lambda^2-1) \\
 2\lambda\sin\alpha & 0 &0 \\
(\lambda^2-1) & 0 & 0
\end{array}
\right]
\label{QuadruX}
\end{equation}
and spherical components by
\begin{equation}
\left\{ \begin{array}{ll}
\displaystyle Q_{22}&\displaystyle= i\sqrt{\frac{15}{2\pi}}\frac{L\lambda\mu \sin\alpha }{\sqrt{1+2\lambda\cos\alpha +\lambda^2}}\\
&\\
\displaystyle Q_{21} &\displaystyle=\sqrt{\frac{15}{8\pi}}\frac{L\mu (1-\lambda^2)}{\sqrt{1+2\lambda\cos\alpha +\lambda^2}} \\
&\\
\displaystyle Q_{20} &\displaystyle=0
\end{array}
\right.
\label{QuadruS}
\end{equation}
For enantiomers, we have $Q_{22}(\mbox{R$_\alpha$-HS})\equiv-Q_{22}(\mbox{S$_\alpha$-HS})$ and  $Q_{21}(\mbox{R$_\alpha$-HS})\equiv Q_{21}(\mbox{S$_\alpha$-HS})$. The magnitude of the quadrupole is \cite{Gray:84}
\begin{equation}
\displaystyle \hat{Q}_2^2=\frac{4\pi}{5}\sum_m \mid Q_{2m}\mid^2=\frac{3}{2}L^2\mu^2\left[\frac{4\lambda^2\sin^2\alpha+(1-\lambda^2)^2}{1+2\lambda\cos\alpha+\lambda^2}\right]
\label{QuadruM}
\end{equation}
Obviously, one has $\hat{Q}_2^2(\mbox{R$_\alpha$-HS})=\hat{Q}_2^2(\mbox{S$_\alpha$-HS})$. In Eq.(\ref{QuadruM}), it might be surprising to find $\hat{Q}_2^2\neq 0$ for $\lambda=0$, since there is only a single dipole in the hard sphere in these cases. This stems from dependence on the choice of the origin in the computation of the quadrupole (the single dipole is shifted by $-L\hat{\bm{u}}$ from the center of the hard sphere - see also ref.\cite{Gray:84}, p.69). This origin dependence highlight the fact that when setting $\lambda=0$, the chiral bidipolar hard sphere model do not reduce to the classical dipolar hard sphere model.  On Figs.(\ref{Fig2}-\ref{Fig3}), we show the  dipole and quadrupole components as functions of the chiral parameter $\chi$ and angle $\alpha$ ; on these representations, the influence of chirality is related to odd or even properties of the function. As shown on Fig.\ref{Fig2}-\ref{Fig3}, a difference between enantiomers R$_\alpha$ and S$_\alpha$ is found only in $Q_{22}$.\\
The spherical components of the octopole are given by
\begin{equation}
\left\{ \begin{array}{ll}
\displaystyle Q_{33}&\displaystyle= 0\\
&\\
\displaystyle Q_{32}&\displaystyle=\frac{1}{2}\sqrt{\frac{105}{8\pi}}L^2\mu (1+2\lambda\cos\alpha +\lambda^2)^{1/2}=\frac{1}{2}\sqrt{\frac{35}{2}}L^2 Q_{10} \\
&\\
\displaystyle Q_{31}&\displaystyle=0\\
&\\
\displaystyle Q_{30} &\displaystyle=-\frac{3}{2}\sqrt{\frac{7}{4\pi}}L^2\mu (1+2\lambda\cos\alpha +\lambda^2)^{1/2}=-\frac{\sqrt{21}}{2}L^2 Q_{10}\\
\end{array}
\right.
\label{OctoS}
\end{equation}
and its magnitude is
\begin{equation}
\displaystyle \hat{Q}_3^2=\frac{4\pi}{7}\sum_m \mid Q_{3m}\mid^2=\frac{33}{8}L^4\mu^2(1+2\lambda\cos\alpha +\lambda^2)=\frac{11\pi}{2}L^4Q_{10}^2
\label{OctoM}
\end{equation}
It is apparent also on Eqs.(\ref{OctoS}-\ref{OctoM}) that the octopole moments are the same for both enantiomers. More generally, by using Eq.(\ref{disdip}) and the definition of the spherical component of multipoles Eq.(\ref{EqA4}), we found
\begin{equation}
\begin{array}{ll} 
\displaystyle \frac{Q_{ll}}{\mu  L^{(l-1)}} = i l \frac{\lambda\sin\alpha}{\sqrt{1+2\lambda\cos\alpha+\lambda^2}}(1+(-1)^l)\mbox{Y}_{ll}(\hat{\bm{e}}_x)
\end{array}
\label{MultiS_ll}
\end{equation}
and, for $0\leq m\leq l-1$, we have
\begin{equation}
\begin{array}{ll} 
\displaystyle \frac{Q_{lm}}{\mu L^{(l-1)}} &\displaystyle =-\sqrt{(l+m+1)(l-m)}\left[\frac{(\lambda^2-(-1)^l)+(1-(-1)^l)\lambda\cos\alpha}{\sqrt{1+2\lambda\cos\alpha+\lambda^2}}\right]\mbox{Y}_{l(m+1)}(\hat{\bm{e}}_x)\\
&\\
&\displaystyle + im \frac{\lambda\sin\alpha}{\sqrt{1+2\lambda\cos\alpha+\lambda^2}}(1+(-1)^l)\mbox{Y}_{lm}(\hat{\bm{e}}_x)
\end{array}
\label{MultiS_lm}
\end{equation}
The value of $\mbox{Y}_{lm}(\hat{\bm{e}}_x)$ may be easily computed, it comes
\begin{equation}
\displaystyle \mbox{Y}_{lm}(\hat{\bm{e}}_x) = \sqrt{\frac{2l+1}{4\pi}} \left|\begin{array}{ll} 
								              \displaystyle (-1)^{(l+m)/2}\frac{[(l-m)!(l+m)!]^{1/2}}{(l-m)!! (l+m)!!} & \mbox{for $(l+m)$ even}\\
								              &\\
								              0  & \mbox{for $(l+m)$ odd}
								              \end{array}
								              \right.
\label{Y_lm_ex}
\end{equation}
therefore, only one of the two contributions in the right handed side of Eq.(\ref{MultiS_lm}) gives $Q_{lm}$. From this computation, we see that all spherical components can be computed only from $Q_{10}$, $Q_{21}$ and $Q_{22}$ ; more precisely, from Eqs.(\ref{MultiS_ll}-\ref{MultiS_lm}), we have 
\begin{equation}
\left\{ \begin{array}{ll} 
\displaystyle \frac{Q_{ll}}{\mu  L^{(l-1)}} &\displaystyle = l(1+(-1)^l)  \sqrt{\frac{2\pi}{15}}\left(\frac{Q_{22}}{\mu L}\right)\mbox{Y}_{ll}(\hat{\bm{e}}_x)\\
&\\
\displaystyle \frac{Q_{lm}}{\mu L^{(l-1)}} &\displaystyle = (1+(-1)^l)\sqrt{\frac{8\pi}{15}}\left[\frac{\sqrt{(l+m+1)(l-m)}}{2}\left(\frac{Q_{21}}{\mu L}\right)\mbox{Y}_{l(m+1)}(\hat{\bm{e}}_x)\right.\\
&\\
&\displaystyle\left.+m\left(\frac{Q_{22}}{\mu L}\right)\mbox{Y}_{lm}(\hat{\bm{e}}_x)\right]\\
&\\
&\displaystyle -\frac{(1-(-1)^l)}{2}\sqrt{(l+m+1)(l-m)}\sqrt{\frac{4\pi}{3}}\left(\frac{Q_{10}}{\mu}\right)\mbox{Y}_{l(m+1)}(\hat{\bm{e}}_x)
\end{array}
\right.
\label{MultiS2_lm}
\end{equation}
Since R$_{(-\alpha)}$-HS $\equiv$ S$_\alpha$-HS, we have
\begin{equation}
\displaystyle Q_{lm}(\mbox{R}_\alpha)-Q_{lm}(\mbox{S}_\alpha)=  2(1+(-1)^l) mL^{(l-2)}\sqrt{\frac{2\pi}{15}}Q_{22}\mbox{Y}_{lm}(\hat{\bm{e}}_x)
\label{Diff_Multi_RS}
\end{equation}
thus, all spherical components of multipoles of enantiomers R$_\alpha$-HS and S$_\alpha$-HS are equal, but those with $l$ and $m$ even and $m\neq 0$. It is worthwhile to note that $Q_{lm}$ is proportional to only one of the three components $Q_{10}$, $Q_{21}$ or $Q_{22}$, because of Eq.(\ref{Y_lm_ex}). Eq.(\ref{MultiS2_lm}) shows that the behavior of all components of multipoles are given (up to a numerical multiplicative factor) by Figs.(\ref{Fig2} -\ref{Fig3}) as functions of the model parameters ; in particular the difference $Q_{lm}(\mbox{R}_\alpha)-Q_{lm}(\mbox{S}_\alpha)$  is given in Fig.\ref{Fig2}(b), as function of $\chi$, and in Fig.\ref{Fig3}(b), as function of $\alpha$.\\
As shown in Eq.(\ref{MultiS2_lm}) all components of the multipoles dependent only on the three components $Q_{10}$, $Q_{21}$ or $Q_{22}$, therefore it is advantageous to define the three reduced geometrical parameters $\chi^*$, $P^*$ and $r^*$ as 
\begin{equation}
\left\{ \begin{array}{ll} 
\displaystyle \chi^* &\displaystyle = \frac{\chi}{L\mu^2}=\lambda\sin\alpha \\
&\\ 
\displaystyle P^*&\displaystyle = \frac{P}{\mu}=(1+2\lambda\cos\alpha+\lambda^2)^{1/2}\\
&\\
\displaystyle r^*&\displaystyle = \frac{r}{L\mu}=(1-\lambda^2)  
\end{array}
\right.
\label{rgeom_para}
\end{equation}

\subsection{Interaction energies between two enantiomers.}

The interaction energy between two enantiomers can be computed in two ways. First, by summing up all interaction energies due to dipole-dipole interactions and/or, second, by using the multipole expansion. In the next section, we give some preliminary numerical results obtained with Monte Carlo Metropolis sampling of the phase space ; in these computations, all dipole-dipole interactions are taken into account explicitely and long ranged interactions are computed with the Ewald method \cite{DeLeeuw:80,Weis:05c}.\\
In the present subsection, we give some analytical results for the interaction energy between two enantiomers by using the multipole expansion when $r\gg\sigma>2L$ and we give also some analytical and numerical results obtained by computing explicitly all dipole-dipole interactions between two enantiomers at contact ({\it i.e.\/} $r=\sigma>2L$).\\
From the multipole expansion, the interaction energy between the multipoles of orders $l_1$ and $l_2$ for two molecules of general shape is given by \cite{Gray:84} 
\begin{equation}
\begin{array}{ll} 
\displaystyle U_{l_1l_2}  = A_{l_1l_2}\sum_{n_1,n_2}\left(\frac{Q_{l_1n_1}Q_{l_2n_2}}{r^{l+1}}\right) &\displaystyle\sum_{m_1,m_2,m}C(l_1 l_2 l;m_1 m_2 m)\\
&\\
&\displaystyle \times D_{m_1n_1}^{l_1}(\omega_1)^* D_{m_2n_2}^{l_2}(\omega_2)^*  \mbox{ Y}_{lm}(\omega)^* 
\end{array}
\label{Multi_l1l2}
\end{equation}
where $l=l_1+l_2$ and $A_{l_1l_2}$ is a numerical constant, $Q_{ln}$ the components of multipoles in the body-fixed frame, $C(l_1 l_2 l;m_1 m_2 m)$ the Clebsch-Gordan coefficients, $D_{mn}^{l}$ are the rotation matrices, the Euler angles $\omega_a\equiv(\phi_a\theta_a\chi_a)$ define the orientation of molecule $a$ and $\omega\equiv\hat{\bm{r}}$.\\
The first contribution to interaction energy is given by the total dipole-total dipole interaction and it is given by
\begin{equation}
\displaystyle U_{11}(a\equiv \mbox{X}_\alpha,b\equiv \mbox{Y}_\alpha) = \frac{1}{r^3}\left[\bm{P}_a\bm{P}_b - 3(\bm{P}_a.\hat{\bm{r}})(\bm{P}_b.\hat{\bm{r}})\right]
\label{U_dipdip}
\end{equation}
where $\mbox{X}_\alpha$, $\mbox{Y}_\alpha\equiv \mbox{R}_\alpha$ or $\mbox{S}_\alpha$. This contribution is not discriminatory between enantiomers (see Eq.(\ref{dipole})) and it is the longest ranged interaction between two chiral dipolar hard spheres. Therefore, many of the properties of this model will coincide with the properties of dipolar hard sphere systems that have been extensively studied by Jean-Jacques Weis in the last decade \cite{Weis:93a,Weis:93b,Levesque:94,Lomba:00,Weis:02,Weis:02a,Tavares:02,Weis:03,Fernaud:03a,Fernaud:03b,Weis:05a,Weis:05b,Weis:05c,Holm:05,Tavares:06,Weis:06,Alvarez:08,Richardi:08}.\\
The discriminatory interaction between enantiomers (Chirodiastaltic interaction \cite{note1}), may be defined by the difference $\Delta U^{(LU)} = U(\mbox{R}_\alpha, \mbox{R}_\alpha)- U(\mbox{R}_\alpha, \mbox{S}_\alpha)$ (where $LU$ stands for $Like-Unlike$), we note also $\Delta U^{(LU)}_{l_1 l_2}=U_{l_1l_2}(\mbox{R}_\alpha,\mbox{R}_\alpha)-U_{l_1l_2}(\mbox{R}_\alpha,\mbox{S}_\alpha)$ as the contribution of multipoles  of orders $l_1$ and $l_2$ to the discriminatory chirodiastaltic interaction. In the following, we choose the molecule at origin to be an R$_\alpha$ enantiomer and all coordinates of vectors and positions are given in its molecule-fixed local frame given by Eq.(\ref{frame}) - this is equivalent to choose $\omega_1\equiv (000)$ in Eq.(\ref{Multi_l1l2}).  From equation (\ref{Diff_Multi_RS}), we have $\Delta U^{(LU)}_{l_1 l_2}\neq 0$ only if $l_2$ even, thus, the first non zero contribution to $\Delta U^{(LU)}$ is given by the dipole-quadrupole interaction ($l_1=1$, $l_2=2$) and after some algebra, it comes
\begin{equation}
\displaystyle \Delta U_{12}^{(LU)}=\left(\frac{L\mu^2}{r^4}\right)\lambda\sin\alpha\left[15\mbox{ M}_2(\phi_2\theta_2\chi_2;\theta\phi)+12\mbox{ M}_1(\phi_2\theta_2\chi_2;\theta\phi)+9\mbox{ M}_0(\theta_2\chi_2;\theta)\right]
\label{DU_12}
\end{equation}
with
\begin{equation}
\left\{ \begin{array}{ll} 
\displaystyle \mbox{M}_2(\phi_2\theta_2\chi_2;\theta\phi)&\displaystyle = \sin^2\theta\cos\theta\left[(1+\cos^2\theta_2)\sin2\chi_2\cos2(\phi_2-\phi) \right.\\
&\\
 &\displaystyle\left. +2\cos\theta_2\cos2\chi_2\sin2(\phi_2-\phi)\right]\\
&\\
\displaystyle \mbox{M}_1(\phi_2\theta_2\chi_2;\theta\phi)&\displaystyle =\sin\theta (5\cos^2\theta -1)\left[\cos2\chi_2\sin(\phi_2-\phi)\right.\\
&\\
&\displaystyle\left. +\cos\theta_2\sin2\chi_2\cos(\phi_2-\phi) \right] \\
&\\
\displaystyle \mbox{M}_0(\theta_2\chi_2;\theta)&\displaystyle = (5\cos^3\theta-3\cos\theta)\sin^2\theta_2\sin2\chi_2
\end{array}
\right.
\label{DU_M12}
\end{equation}
where $\omega_2\equiv(\phi_2 \theta_2 \chi_2)$ and $\omega=(\phi\theta)$ are respectively the orientations of the second enantiomer and of the bond vector $\hat{\bm{r}}$ in the molecule-fixed local frame of the first enantiomer.\\
The quadrupole-quadrupole interaction contributes also to $\Delta U^{(LU)}$ ; $\Delta U^{(LU)}_{22}$ may be written as
\begin{equation}
\begin{array}{ll}
\displaystyle \Delta U_{22}^{(LU)}&\displaystyle = -\frac{3}{2}\left(\frac{L^2\mu^2}{r^5}\right)\mbox{ f}_{22}(\lambda,\alpha)\left[A_0+20 A_1+30\sqrt{2}A_2+140 A_3+35 A_4\right]\\
&\\
&\displaystyle -6\left(\frac{L^2\mu^2}{r^5}\right)\mbox{ f}_{21}(\lambda,\alpha)\left[B_0+\frac{5}{\sqrt{2}}B_1-5\sqrt{2}B_2-\frac{35}{4}B_3\right]
\end{array}
\label{DU_22}
\end{equation}
with the functions f$_{22}$ and f$_{21}$ describing the dependence of $\Delta U_{22}^{(LU)}$ on the model parameters ; these two functions are given by
\begin{figure}
\centerline{\includegraphics[width=5in]{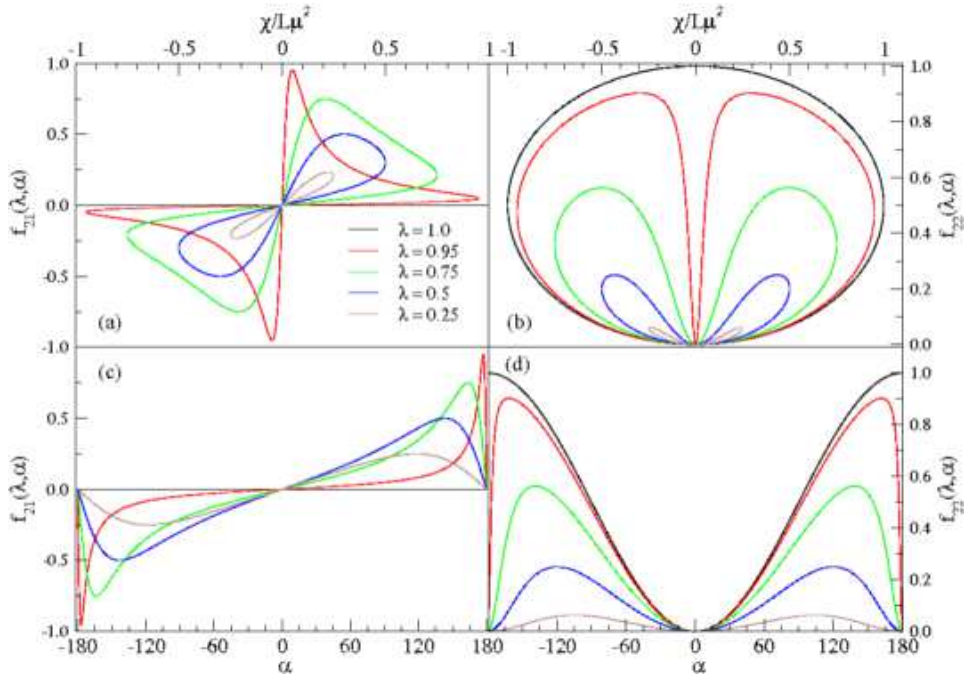}}
\caption{{\bf } Representation of the two functions f$_{21}$ and f$_{22}$ as function of the reduced chiral parameter $\chi^*=\chi/L\mu^2$ and as functions of $\alpha$.}
\label{Fig4}
\end{figure}
\begin{equation}
\left\{ \begin{array}{ll} 
\displaystyle \mbox{ f}_{22}(\lambda,\alpha)&\displaystyle = \frac{\lambda^2\sin^2\alpha}{(1+2\lambda\cos\alpha+\lambda^2)}=\frac{\chi^{*2}}{P^{*2}}\\
&\\
\displaystyle \mbox{ f}_{21}(\lambda,\alpha)&\displaystyle = \frac{(1-\lambda^2)\lambda\sin\alpha}{(1+2\lambda\cos\alpha+\lambda^2)}=\frac{r^*\chi^{*}}{P^{*2}}
\end{array}
\right.
\label{DU22_f}
\end{equation}
In Figs.\ref{Fig4} (a,b), we represent these two functions versus the chiral parameter $\chi$ and, in Figs.\ref{Fig4} (c,d) as functions of $\alpha$.\\
The functions $A$ and $B$ depend on orientations $\omega_2$ and $\omega$. More precisely, we have
\begin{equation}
\left\{ \begin{array}{ll} 
\displaystyle A_0 &\displaystyle = (35\cos^4\theta-30\cos^2\theta+3)[(1+\cos^2\theta_2)\sin2\chi_2\sin2\phi_2-2\cos\theta_2\cos2\chi_2\cos2\phi_2]\\
&\\
\displaystyle A_1 &\displaystyle = \sin\theta \mbox{ }(7\cos^3\theta-3\cos\theta)\sin\theta_2[\cos2\chi_2\cos(\phi_2+\phi)-\cos\theta_2\sin2\chi_2\sin(\phi_2+\phi)] \\
&\\
\displaystyle A_2 &\displaystyle =  \sin^2\theta\mbox{ } (7\cos^2\theta-1)\sin2\phi \sin^2\theta_2\sin2\chi_2 \\
&\\
\displaystyle A_3 &\displaystyle = \sin^3\theta\cos\theta\sin\theta_2 [ \cos2\chi_2 \cos(\phi_2-3\phi)-\cos\theta_2\sin2\chi_2\sin(\phi_2-3\phi)] \\
&\\
\displaystyle A_4 &\displaystyle = \sin^4\theta[2\cos\theta_2\cos2\chi_2\cos(2\phi_2-4\phi)-(1+\cos^2\theta_2)\sin2\chi_2\sin(2\phi_2-4\phi)]
\end{array}
\right.
\label{DU22_A}
\end{equation}
and
\begin{equation}
\left\{ \begin{array}{ll} 
\displaystyle B_0 &\displaystyle =(35\cos^4\theta-30\cos^2\theta+3)\sin\theta_2[\cos2\chi_2\cos\phi_2+\cos\theta_2\cos2\chi_2\cos\phi_2] \\
&\\
\displaystyle B_1 &\displaystyle = \sin\theta \mbox{ }(7\cos^3\theta-3\cos\theta) \left[(1+\cos^2\theta_2)\sin2\chi_2\sin(2\phi_2-\phi)\right.\\
&\\
&\displaystyle +2\cos\theta_2\cos2\chi_2\cos(2\phi_2-\phi)-\frac{3}{2\sqrt{2}}\sin^2\theta_2\sin2\chi_2\cos\phi]\\
&\\
\displaystyle B_2 &\displaystyle = \sin^2\theta\mbox{ } (7\cos^2\theta-1)\sin\theta_2 [ \cos2\chi_2\sin(\phi_2-2\phi)+\cos\theta_2\sin2\chi_2\sin(\phi_2-2\phi)] \\
&\\
\displaystyle B_3 &\displaystyle = \sin^3\theta\cos\theta [(1+\cos^2\theta_2)\sin2\chi_2\sin(2\phi_2-3\phi)-2\cos\theta_2\cos2\chi_2\cos(2\phi_2-3\phi)]
\end{array}
\right.
\label{DU22_B}
\end{equation}
Obviously, if the distance between two enantiomers is too small the multipole expansion ceases to be accurate, however, the interaction energy between enantiomers may still be computed by summing all dipole-dipole interactions. The total interaction energy between two enantiomers $a$ and $b$ has thus to be written as
\begin{figure}
\centerline{\includegraphics[width=4in]{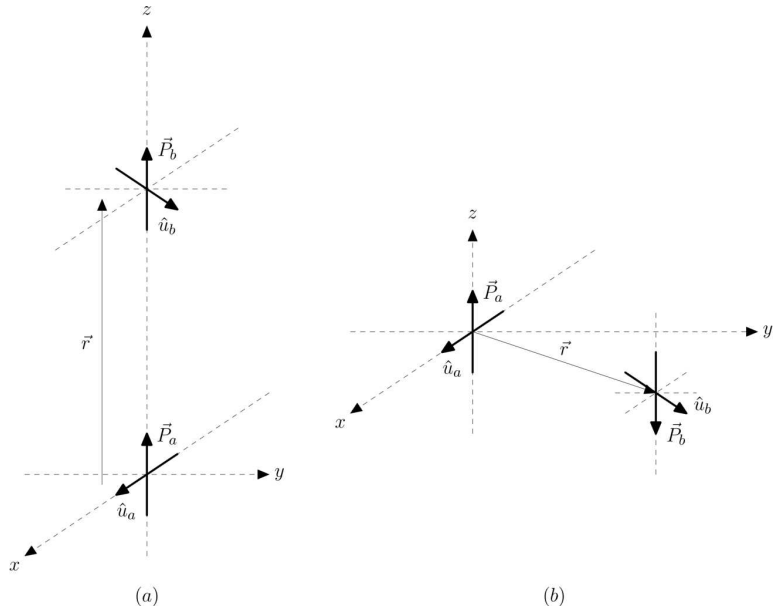}}
\caption{{\bf} Configurations of particular interest in dipolar system. (a) Colinear aligned total dipole moments, these kind of configurations between two dipolar particle is responsible for chains formation at low temperature and low density : for this configuration, one has $\theta=0$ and $\theta_2=0$. (b) Antiparallel total dipole moment : for this configuration, one has $\theta=\pi/2$ and $\theta_2=\pi$.}
\label{Fig5}
\end{figure}
\begin{equation}
E^{(a,b)}=E^{(1a,1b)}+E^{(1a,2b)}+E^{(2a,1b)}+E^{(2a,2b)}
\label{Tot_ab_int}
\end{equation}
where $E^{(ia,jb)}$ denotes the interaction energy between the dipole $i$ of enantiomer $a$ with dipole $j$ of enantiomer $b$ ; it is given by
\begin{equation}
\displaystyle E^{(ia,jb)}=\frac{1}{\mid \bm{r}^{ij}_{ab}\mid^3}\left[\bm{\mu}_i^{(a)}.\bm{\mu}_j^{(b)}-3(\bm{\mu}_i^{(a)}.\hat{\bm{r}}^{ij}_{ab})(\bm{\mu}_j^{(b)}.\hat{\bm{r}}^{ij}_{ab})\right]
\label{iajb_int}
\end{equation}
and the distance vectors between point dipoles are 
\begin{figure}
\centerline{\includegraphics[width=6in]{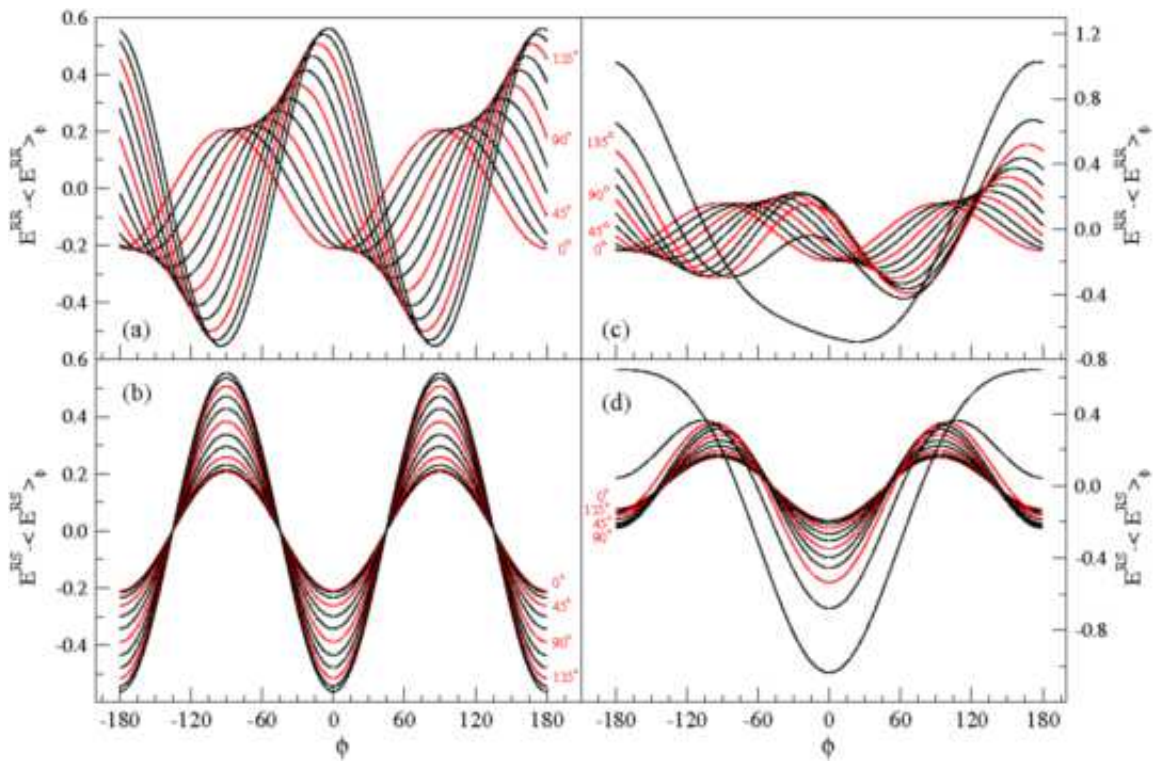}}
\caption{{\bf } Interaction energies for two enantiomers in configurations given by Fig.\ref{Fig5} (a) at contact $z=\sigma$. The angle $\phi$ is defined by $\cos\phi = \hat{\bm{u}}_a.\hat{\bm{u}}_b$. In all computations reported in these figures, $\mu=1.0$, $L=0.25\sigma$ and the curves correspond to an increase in the value of $\alpha$ by $15^o$ ; the red lines correspond to $\alpha=0^o$, $45^o$, $90^o$ and $135^o$, these values are indicated at edge of curves in figures. The energy $<E^{XY}>_\phi$ is the average energy over the orientation $\hat{\bm{u}}_b$ of enantiomers $Y$. (a) $X$ and $Y\equiv\mbox{R}_\alpha$ and $\lambda=1.0$ ; (b) $Y\equiv\mbox{S}_\alpha$ and $\lambda=1.0$ ; (c) $X$ and $Y\equiv\mbox{R}_\alpha$ and $\lambda=0.75$ and (d) $Y\equiv\mbox{S}_\alpha$ and $\lambda=0.75$.}
\label{Fig6}
\end{figure}
\begin{table}
 \tbl{ Average energy over the orientation $\hat{\bm{u}}_b$ of enantiomers $Y$ and location $\phi_m$ of the orientation of $\hat{\bm{u}}_b$, for which the minimun energy is reached, when $Y\equiv\mbox{R}_\alpha$ for configurations given by Fig.\ref{Fig5} (a) at contact $z=\sigma$.}
{\begin{tabular}{@{}lcccc}\toprule
	        & $<E^{XY}>_{\phi}$ &                            & $\phi_m$ & \\
$\alpha$  & $\lambda=1$ & $\lambda=0.75$ & $\lambda=1$ & $\lambda=0.75$ \\
\colrule
 0$^o$      &    -5.794        &        -4.436          & 0$^o$             & 0$^o$ \\
 15$^o$    &    -5.695        &        -4.362          & 11.2$^o$        & 10.4$^o$ \\
 30$^o$    &    -5.506        &        -4.145         & 22.4$^o$        & 20.8$^o$ \\
 45$^o$    &    -4.945        &        -3.800          & 33$^o$           & 31$^o$ \\
 60$^o$    &    -4.345        &        -3.350         & 43$^o$           & 40.2$^o$ \\
 75$^o$    &    -3.647        &        -2.825         & 51.6$^o$        & 48.4$^o$ \\
 90$^o$    &    -2.897        &        -2.263         & 59.2$^o$        & 55.4$^o$ \\
105$^o$   &    -2.147        &        -1.701         & 65.8$^o$        & 61$^o$ \\
120$^o$   &    -1.448        &        -1.177         & 71.4$^o$        & 65.2$^o$ \\
135$^o$   &    -0.848        &        -0.727         & 76.6$^o$        & 67$^o$ \\
150$^o$   &    -0.388        &        -0.381         & 81.2$^o$        & 62.8$^o$ \\
165$^o$   &    -0.098        &        -0.164         & 85.6$^o$        & 11$^o$ \\
\botrule
\end{tabular}}
\label{AverageTable}
\end{table}
\begin{equation}
\left\{ \begin{array}{ll} 
\displaystyle \bm{r}^{11}_{ab}&\displaystyle = \bm{r}+L(\hat{\bm{u}}_a-\hat{\bm{u}}_b)\\
&\\
\displaystyle \bm{r}^{12}_{ab}&\displaystyle = \bm{r}+L(\hat{\bm{u}}_a+\hat{\bm{u}}_b)\\
&\\
\displaystyle \bm{r}^{21}_{ab}&\displaystyle = \bm{r}-L(\hat{\bm{u}}_a+\hat{\bm{u}}_b)\\
&\\
\displaystyle \bm{r}^{22}_{ab}&\displaystyle = \bm{r}-L(\hat{\bm{u}}_a-\hat{\bm{u}}_b)
\end{array}
\right.
\label{dist_iajb}
\end{equation}
Even for simple relative configurations, as the ones shown on Fig.\ref{Fig5}(a-b), the total interaction energy between two enantiomers is a bit complicated when computed by summing all dipole-dipole interactions. For instance, the interaction energy between two $\mbox{R}_\alpha$ enantiomers  in a configuration given by Fig.\ref{Fig5}(a) is
\begin{equation}
\begin{array}{ll} 
\displaystyle E^{(a,b)}&\displaystyle =\frac{\mu^2}{\mid z\mid^3}\\ 
&\\
&\displaystyle\times\left\{\frac{1}{(1+2\left(\frac{L^2}{z^2}\right)(1-\cos\phi))^{5/2}}\left[\left(P^{*2}+\frac{r^{*2}}{P^{*2}}\right)\left(\left(\frac{L}{z}\right)^2(1-\cos\phi)-1\right)\right.\right.\\
&\\
&\displaystyle \left. -6\left(\frac{L}{z}\right)\chi^*\sin\phi+2\left(\frac{\chi^{*2}}{P^{*2}}\right)\left(\cos\phi-\left(\frac{L}{z}\right)^2(2-2\cos\phi+\sin^2\phi)\right)\right]\\
&\\
&\displaystyle +\frac{1}{(1+2\left(\frac{L^2}{z^2}\right)(1+\cos\phi))^{5/2}}\left[\left(P^{*2}-\frac{r^{*2}}{P^{*2}}\right)\left(\left(\frac{L}{z}\right)^2(1+\cos\phi)-1\right)\right.\\
&\\
&\displaystyle \left.\left. +6\left(\frac{L}{z}\right)\chi^*\sin\phi-2\left(\frac{\chi^{*2}}{P^{*2}}\right)\left(\cos\phi-\left(\frac{L}{z}\right)^2(2+2\cos\phi+\sin^2\phi)\right)\right]\right\}
\end{array}
\label{Eab_5a}
\end{equation}
where $\cos\phi=\hat{\bm{u}}_a.\hat{\bm{u}}_b$ and $(\chi^*,P^*,r^*)$ given by Eq.(\ref{rgeom_para}). The chirodiastaltic interaction between enantiomers $\mbox{R}_\alpha$ and  $\mbox{S}_\alpha$, computed by summing the four dipole-dipole interaction energies, for the configuration given on Fig.\ref{Fig5}(a) is
\begin{figure}
\centerline{\includegraphics[width=6in]{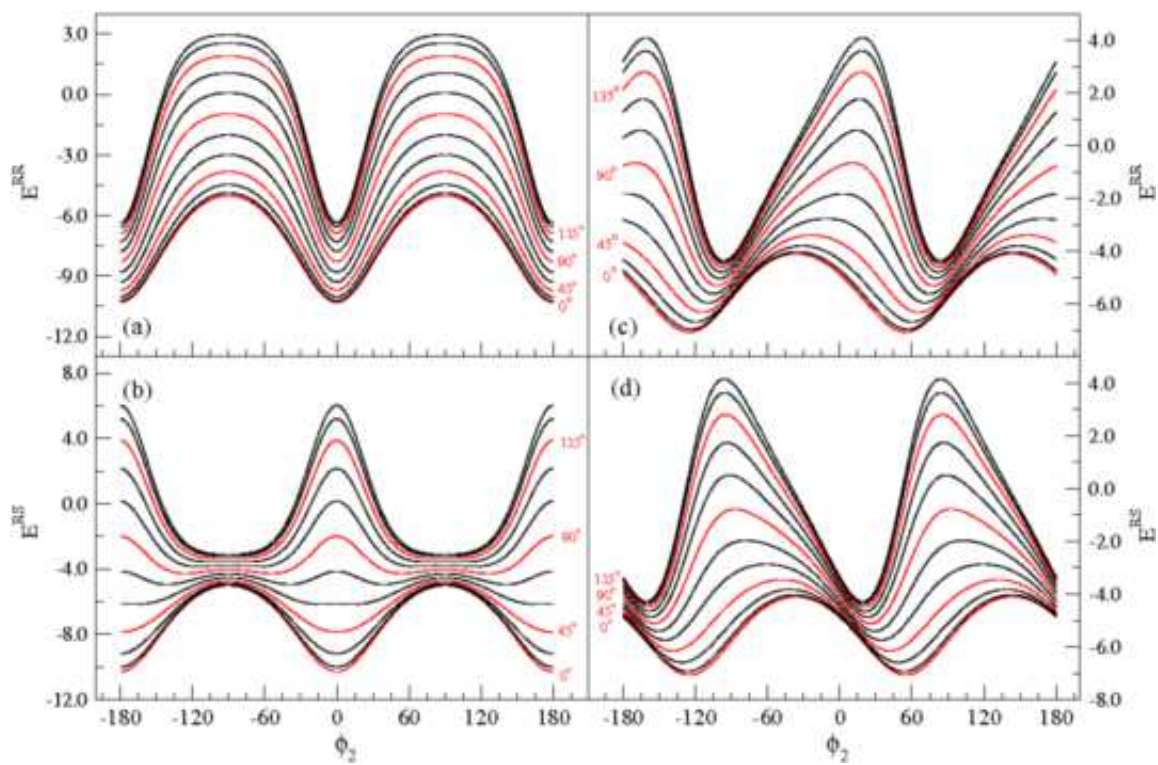}}
\caption{{\bf } Interaction energies for two enantiomers in configurations given by Fig.\ref{Fig5} (b) at contact $r=\sigma$ for $\lambda=1$. In all computation, $\mu=1.0$, $L=0.25\sigma$ and notations are the same as in Fig.\ref{Fig6}. For the configuration given by Fig.\ref{Fig5} (b), the angle $\phi$ is defined by $\cos\phi= \hat{\bm{u}}_a.\hat{\bm{r}}$ and the angle $\phi_2$ by $\cos\phi_2 = \hat{\bm{u}}_a.\hat{\bm{u}}_b$. (a-b) : $\phi=0^o$  and (c-d) : $\phi=45^o$.}
\label{Fig7}
\end{figure}
\begin{figure}
\centerline{\includegraphics[width=6in]{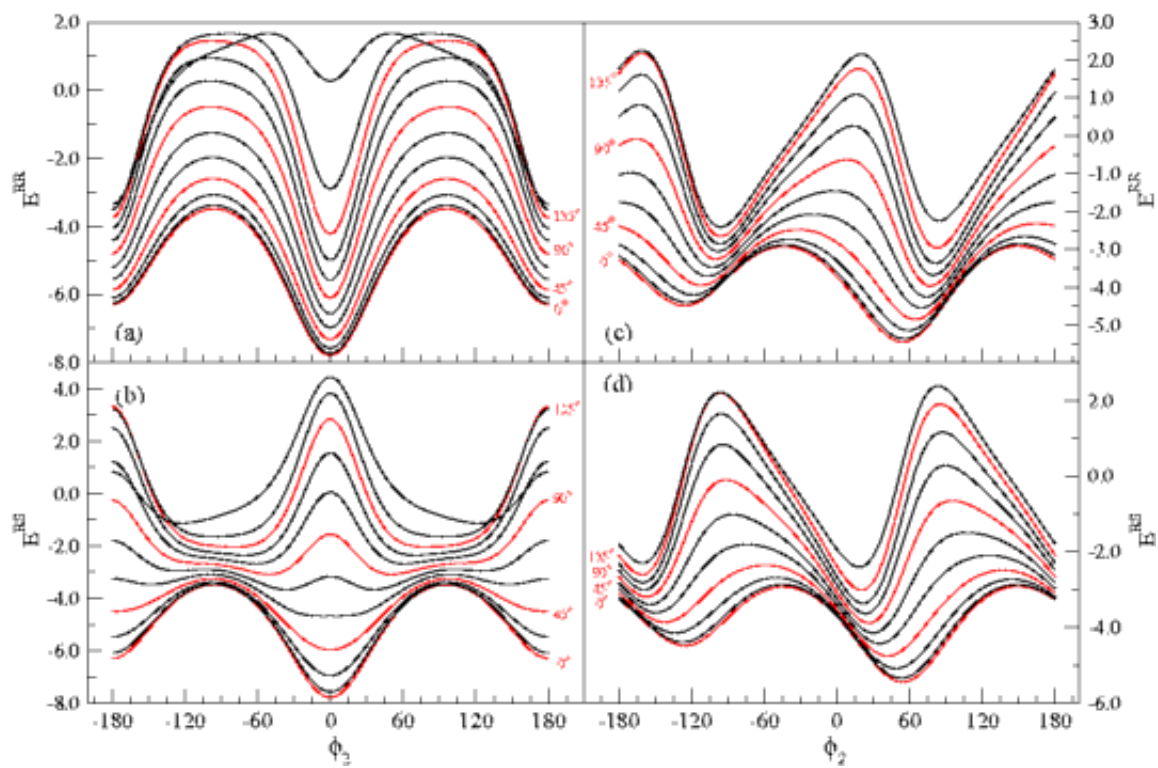}}
\caption{{\bf } Interaction energies between two enantiomers in configurations given by Fig.\ref{Fig5} (b) at contact $r=\sigma$ for $\lambda=0.75$. Notations are the same as for Fig.\ref{Fig7}.}
\label{Fig8}
\end{figure}
\begin{equation}
\begin{array}{ll} 
\displaystyle\Delta E^{(LU)}&\displaystyle = E^{(\mbox{R}_\alpha,\mbox{R}_\alpha)}-E^{(\mbox{R}_\alpha,\mbox{S}_\alpha)}=\frac{\mu^2}{\mid z\mid^3}\\
&\\
&\displaystyle\times\left\{\frac{1}{(1+2\left(\frac{L^2}{z^2}\right)(1-\cos\phi))^{5/2}}\left[4\frac{\chi^{*2}}{P^{*2}}\cos\phi-6\left(\frac{L}{z}\right)\chi^*\sin\phi\right.\right.\\
&\\
&\displaystyle\left.-4\left(\frac{L^2}{z^2}\right)\frac{\chi^{*2}}{P^{*2}}(\sin^2\phi+2-2\cos\phi)\right]\\
&\\
&\displaystyle+\frac{1}{(1+2\left(\frac{L^2}{z^2}\right)(1+\cos\phi))^{5/2}}\left[-4\frac{\chi^{*2}}{P^{*2}}\cos\phi+6\left(\frac{L}{z}\right)\chi^*\sin\phi\right.\\
&\\
&\displaystyle\left.\left.-4\left(\frac{L^2}{z^2}\right)\frac{\chi^{*2}}{P^{*2}}(\sin^2\phi+2+2\cos\phi)\right]\right\}
\end{array}
\label{diff_5a}
\end{equation}
For $\mid z\mid \gg L$, a Taylor expansion of Eq.(\ref{diff_5a}) gives
\begin{equation}
\begin{array}{ll} 
\displaystyle\Delta E^{(LU)}&\displaystyle = \frac{\mu^2}{\mid z\mid^3}\left[24\frac{\chi^{*2}}{P^{*2}}\left(\frac{L^2}{z^2}\right)\sin2\phi-60\chi^*\left(\frac{L^3}{z^3}\right)\sin\phi+o\left(\frac{L^4}{z^4}\right)\right]\\
&\\
&\displaystyle =  \Delta U_{22}^{(LU)}(\theta_2=0,\theta=0)+ \Delta U_{32}^{(LU)}(\theta_2=0,\theta=0)+o\left(\frac{L^4\mu^2}{z^7}\right)
\end{array}
\label{diff_5a_z}
\end{equation}
where we have set $\phi=\phi_2+\chi_2$ in Eqs.(23-24) to obtain $\Delta U_{22}^{(LU)}$. Substracting Eq.(\ref{diff_5a_z}) from Eq.(\ref{diff_5a}), may allow to check the degree of accuracy of the multipole expansion for the configuration of Fig.\ref{Fig5} (a).\\
On Figs.\ref{Fig6}, we represent interaction energies for enantiomers in the colinear aligned total dipole moment configuration (Fig.\ref{Fig5}-a). These energies have been computed by summing all dipole interactions between the two enantiomers. To allow simple comparison between interaction energies, for different values of $\alpha$, we have substracted to all curves the average, over the orientation $\hat{\bm{u}}_b$, of the interaction energy. For all $\lambda$, a minimum in the interaction energy between two different enantiomers is always obtained for $\phi=0$ ($\hat{\bm{u}}_a$ and $\hat{\bm{u}}_b$ parallel) ; the anti-parallel configuration of molecular axis $\hat{\bm{u}}_a$ and $\hat{\bm{u}}_b$ ($\phi=\pi$) is a relative minimum for the interaction energy for all $\lambda<1$ (see Fig.\ref{Fig6}(b-d) - if $\lambda=1$, the configurations for $\phi=0$ and $\phi=\pi$ have the same energy) ; one should note also that these minima are relative, because the configurations like in Fig.\ref{Fig5} (a) do not permit to reach the smallest distance between two point dipoles. The minimum interaction energy between two enantiomers of same kind is not obtained for parallel or anti-parallel configurations of the molecular axis. The value $\phi_m$ of the angle for which a minimum interaction energy is obtained, depends on parameters $\lambda$ and $\alpha$ (see Fig.\ref{Fig6}(a-c)), several values are given in Table \ref{AverageTable} ; average values of $<E^{XY}>_{\phi}$ are also reported in this Table. Therefore, in homochiral systems (systems that are composed with exclusively one enantiomer) with a large enough total dipole moment and in thermodynamical conditions where dipolar systems form chains \cite{Weis:93a,Levesque:94}, we may observe formation of chains too, and the molecular axis of the enantiomers may have helical arrangement. Addition of  chiral enantiomers in the system will reduce the range of the helical arrangement in the chains. For $\lambda=1$, the minimum at $\phi=\phi_m$ has the same value than the one found at $\phi=\phi_m-\pi$, therefore {\it right} and {\it left-handed} helical arrangement of the molecular axis in the chains will compete in homochiral systems.\\
For the configuration given in  Fig.\ref{Fig5}(b) and for $r\gg L$ ($\theta=\pi/2$ and $\theta_2=\pi$), the multipole-multipole chirodiastaltic interaction is found as
\begin{figure}
\centerline{\includegraphics[width=4.5in]{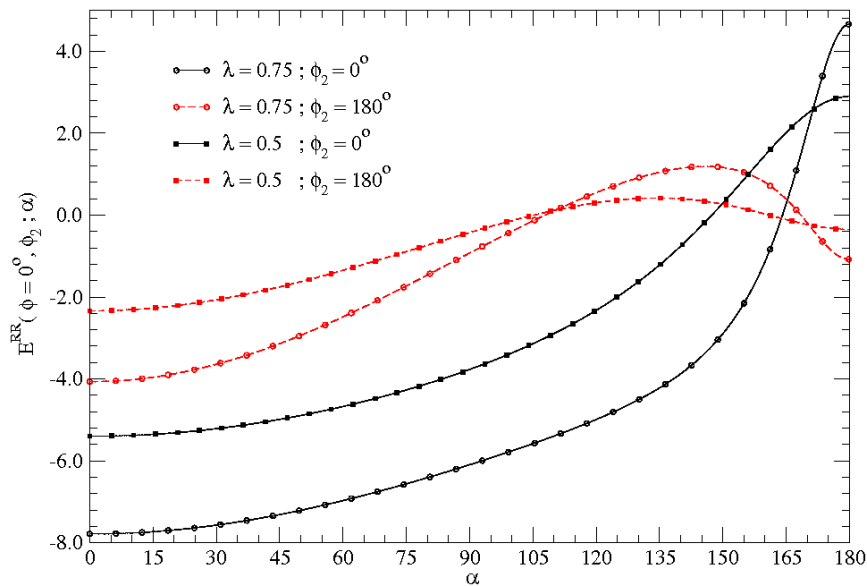}}
\caption{{\bf } Relative minimum of interaction energies between two enantiomers $\mbox{R}_\alpha$ in antiparallel total dipole configurations with $\phi=0$ as functions of $\alpha$. Open circles correspond to $\lambda=0.75$ and filled squares to $\lambda=0.5$, the black solid lines are for $\phi_2=0$ and the red dashed lines for $\phi_2=\pi$. For $\lambda=0.75$, one has $\alpha_m\simeq 166^o$ and for $\lambda=0.5$, $\alpha_m\simeq 150^o$.}
\label{Fig9}
\end{figure}
\begin{equation}
\begin{array}{ll} 
\displaystyle\Delta E^{(LU)}&\displaystyle =-3\frac{\mu^2}{r^3}\left[4\left(\frac{L}{r}\right)\chi^*\sin(\phi_2-\phi-2\chi_2)\right.\\
&\\
&\displaystyle\left.+\left(\frac{L^2}{r^2}\right)\frac{\chi^{*2}}{P^{*2}}\left(3\cos 2(\phi_2-\chi_2)-35\cos 2(\phi_2-\chi_2-2\phi)\right)\right]+o\left(\frac{L^3\mu^2}{z^6}\right)\\
&\\
&\displaystyle =  \Delta U_{12}^{(LU)}(\theta_2=\pi,\theta=\frac{\pi}{2})+ \Delta U_{22}^{(LU)}(\theta_2=\pi,\theta=\frac{\pi}{2})+o\left(\frac{L^3\mu^2}{z^6}\right)
\end{array}
\label{diff_5b_r}
\end{equation}
On Figs.\ref{Fig7} and \ref{Fig8}, we represent interaction energies for enantiomers in the antiparallel total dipole configurations (Fig.\ref{Fig5}-b) for $\phi=0$ and $\pi/4$. On Figs.\ref{Fig7} (a,c) and \ref{Fig8} (a,c), we represent interactions energies between two $\mbox{R}_\alpha$ enantiomers for $\lambda =1$ and $0.75$ and on 
Figs.\ref{Fig7} (b,d) and \ref{Fig8} (b,d) we represent interactions energies between enantiomers $\mbox{R}_\alpha$ and $\mbox{S}_\alpha$. The coupling between $\phi$ and $\phi_2$ renders quite complicated the energy landscape of antiparallel configurations. The relative minima for antiparallel configurations (Fig.\ref{Fig5}-b) are smaller than the minima found for colinear configurations (Fig.\ref{Fig5}-a), therefore the colinear and antiparallel configurations are in drastic competition by comparison to dipolar hard sphere systems.\\
For two identical enantiomers, when $\phi=n \pi/2$ (with $n=-1,0,1$ or $2$), the location of minimum interaction energies are independent on $\alpha$. For instance, if $\phi=0$, the relative minima for two $\mbox{R}_\alpha$ enantiomers are always obtained for $\phi_2=0$ or $\pi$ ; if $\lambda=1$, these two minima have the same value, but if $\lambda < 1$ the two minima are not equivalent and their value dependent on $\alpha$. On Fig.\ref{Fig9}, we give the value of the minimum energy for $\phi=0$ and $\phi_2=0$ or $\pi$ for $\lambda=0.75$ and $0.5$ ; as shown on this figure for values of $\alpha> \alpha_m(\lambda)$ the configuration $\phi_2=\pi$ is more stable than the configuration $\phi_2=0$. The amplitude of variation of interaction energies with $\phi_2$ are larger for antiparallel configurations than variation with $\phi$ for aligned colinear configurations.\\     
The locations of relative minimum interaction energies between enantiomers $\mbox{R}_\alpha$ and $\mbox{S}_\alpha$ depend not only on $\phi$ and $\phi_2$, but also on $\alpha$. For instance on Fig.\ref{Fig7}(b) ($\phi=0$) and for $\alpha=5\pi/12$ ($75^o$), we found four different minima located at $\phi_2\simeq\pm141^o$ and $\pm 39^o$.\\
The computations of interaction energies between two chiral dipolar hard spheres show that a lot of different configurations may be in competition, even for simple configurations as the ones shown on Figs.\ref{Fig5}. In the next section, we give some preliminary results obtained with Monte Carlo simulations for several cases.  

\section{\label{sec:level3} Preliminary Monte Carlo results }

As shown in the previous section, interaction energies between two enantiomers as function of their relative orientation may have a lot of relative minima ; therefore, despite the apparent simplicity of the model, the thermodynamical properties chiral dipolar hard sphere systems are quite complicated. In particular, because the energy landscape of the interaction between two enantiomers shows large variations, one may predict that frustration phenomena will play an important role in phases at low densities, low temperatures and also in solid phases.\\
We define the reduced quantities as follow : for volume of the simulation box $V^*=V/\sigma^3$, density $\rho^*=\rho\sigma^3=N\sigma^3/V$, pressure $p^*=p\sigma^2/kT$ and dipole moment $\mu^*=(\mu/kT\sigma^3)^{1/2}$. For a given model of enantiomer, defined by $(\lambda,\alpha)$ and $L$, the reduced total dipole moment $P^*$ of the molecule is related to $\mu^*$ according to Eq.(\ref{dipole}). In all computations, we have chosen $L^*=L/\sigma=1/4$ as in the numerical computations done in the previous section (see Figs.\ref{Fig6}-\ref{Fig8}). At the end of this section, a short discussion on the influence of $L$ on structures is presented ; this influence is illustrated on two cases : $L^*=0.025$ and $L^*=0.425$. For notational convenience the asterisks will be drop in the following. \\
Mixtures of enantiomers R$_\alpha$-HS and S$_\alpha$-HS are also of interest, we define densities $\rho_{\mbox{\tiny R}}=N_{\mbox{\tiny R}}/V$ and $\rho_{\mbox{\tiny S}}=N_{\mbox{\tiny S}}/V$ respectively for enantiomers R$_\alpha$-HS and S$_\alpha$-HS and the ratio $X=\rho_{\mbox{\tiny S}}/\rho_{\mbox{\tiny R}}=N_{\mbox{\tiny S}}/N_{\mbox{\tiny R}}$. The ratio $X$ is related to the {\it enantiomeric excess\/} ($ee$) in enantiomer R$_\alpha$-HS by $ee=1-X$. In the following, we have considered only two mixtures of enantiomers : homochiral R$_\alpha$ systems ($X=0$) and racemic R$_\alpha$-S$_\alpha$ systems ($X=1$).\\
Monte Carlo simulations have been performed in canonical $(NVT)$ and isobaric $(NPT)$ ensembles with system size $N=N_{\mbox{\tiny R}}+N_{\mbox{\tiny S}}= 512$ or $1000$. Periodic boundary conditions are used and dipolar interaction energies between two enantiomers is computed with Eqs.(\ref{Tot_ab_int}-\ref{dist_iajb}) where long ranged contributions are taken into account by using the Ewald method \cite{Weis:05c}.\\
From a technical point of view, as a consequence of the numerous number of relative minima in interaction energies, the convergence of the Monte Carlo algorithm and the sampling of the phase space will require a large number of MC cycles (one MC cycle corresponds, on average, to one trial move per particle for $NVT$ simulations ; for $NPT$ simulation, an additional trial move of the volme of the box is done every MC-cycle). For most results presented in this section, between $5\times 10^4$ and $10^5$ MC cycles have been achieved for the relaxation and convergence from an initial condition and averages have been accumulated over about $10^5$ and $2\times 10^5$ MC cycles, respectively for systems with 1000 and 512 particles.\\
Periodic boundary conditions are used and interaction energies are computed with the Ewald method \cite{Weis:05c} by summing all dipole-dipole interactions between enantiomers. The average energy $U$ is computed as
\begin{table}
 \tbl{Average energies and eigenvalues $S_+$, $S_0$ and $S_-$ for homochiral-$\mbox{R}_\alpha$ and racemic systems in external electric field $\bm{E} =E\hat{\bm{e}}_z$ ($E=10$) obtained in canonical $(NVT)$ Monte Carlo simulations. The numbers in brackets give the accuracy on the last digit of the averages. For all systems  $N=512$, $\mu=1.0$ and $\rho=0.05$ (or $\rho=0.5$). The parameters $\alpha$ and $\lambda$ define the molecular structure of enantiomers as given by Eq.(\ref{Dipole1}).}
{\begin{tabular}{@{}lcccccccccc}\toprule
	        & Homochiral  &                     &            &            &            & Racemic      &                    &             &            &      \\
$\alpha$  & $\beta U/N$ & $\beta V/N$ & $S_+$ & $S_0$ & $S_-$ & $\beta U/N$ & $\beta V/N$ & $S_+$ & $S_0$ & $S_-$\\
\colrule
$\lambda=1$ &    &                     &            &            &            &      &                    &             &            &      \\
 &    &                     &            &            &            &      &                    &             &            &      \\
$20^o$   & -5.3(2)   & -18.83(5) & 0.25(1) & 0.18(1) & -0.442(4) & -5.3(3)   & -18.85(6) & 0.27(2) & 0.18(2) & -0.442(4) \\
$35^o$   & -4.2(2)   & -18.18(6) & 0.24(1) & 0.19(1) & -0.434(4) & -4.1(3)   & -18.19(6) & 0.25(2) & 0.18(2) & -0.435(4) \\
$45^o$   & -3.0(3)   & -17.55(6) & 0.24(1) & 0.19(1) & -0.429(4) & -2.9(2)   & -17.54(6) & 0.25(2) & 0.18(2) & -0.429(4) \\
$55^o$   & -2.2(1)   & -16.79(6) & 0.25(1) & 0.18(1) & -0.424(4) & -2.2(1)   & -16.79(6) & 0.24(2) & 0.19(2) & -0.424(4) \\
$60^o$   & -1.9(1)   & -16.37(6) & 0.24(1) & 0.19(1) & -0.423(5) & -1.9(1)   & -16.37(6) & 0.24(2) & 0.18(2) & -0.423(4) \\
$90^o$   & -0.83(5) & -13.17(5) & 0.23(1) & 0.17(1) & -0.405(6) & -0.82(5) & -13.17(5) & 0.23(2) & 0.18(2) & -0.404(6) \\
$120^o$ & -0.37(4) & -9.02(5)   & 0.21(1) & 0.16(1) & -0.37(1)   & -0.35(3) & -9.01(5) & 0.21(2) & 0.16(1) & -0.37(1) \\ 
$150^o$ & -0.21(3) & -4.19(5)   & 0.16(2) & 0.11(2) & -0.27(1)   & -0.18(3) & -4.19(4) & 0.16(1) & 0.11(1) & -0.27(1) \\ 
\colrule
$\lambda=0.95$      &    &                     &            &            &            &      &                    &             &            &      \\
 &    &                     &            &            &            &      &                    &             &            &      \\
$170^o$                  &  -0.11(2) & -0.88(3) & 0.06(2) & 0.02(1) & -0.08(2) & -0.10(2) & -0.88(3) & 0.07(1) & 0.02(1) & -0.09(2) \\
$^{\rm a}$$170^o$ &  -1.50(6) & -0.93(3) &  0.08(1) & 0.03(1) & -0.10(2) & -1.44(6) & -0.91(3) & 0.07(1) & 0.03(1) &-0.10(1) \\
\colrule
$\lambda=0.75$      &    &                     &            &            &            &      &                    &             &            &      \\
 &    &                     &            &            &            &      &                    &             &            &      \\
$152^o$                  & -0.09(2) & -3.89(5) & 0.15(1) & 0.10(2) & -0.26(1) & -0.09(2) & -3.90(4) & 0.16(2) & 0.10(2) & -0.26(1) \\
$^{\rm a}$$152^o$ & -1.26(5) & -3.96(4) & 0.16(2) & 0.11(2) & -0.28(1)  & -1.21(5) & -3.94(5) & 0.16(2) & 0.11(2) & -0.27(1) \\
\botrule
\end{tabular}}
\tabnote{$^{\rm a}$ For these computations, the density is $\rho=0.5$.}
\label{Table_MC_champ}
\end{table}
\begin{figure}
\centerline{\includegraphics[angle=270,width=5.in]{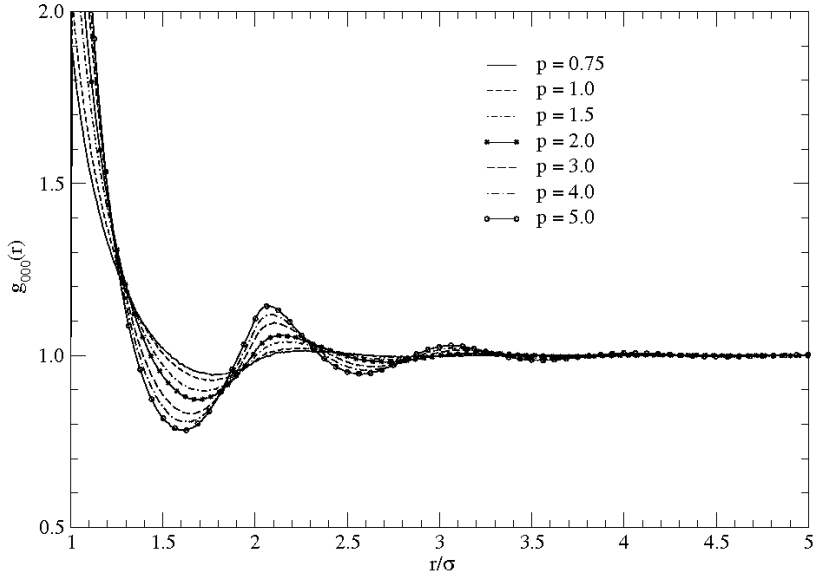}}
\caption{{\bf } Representation of $g_{000}(r)$ for several value of the pressure obtained by Monte-Carlo sampling of the isobaric $(NPT)$ ensemble of homochiral-$\mbox{R}_\alpha$ systems for $N=1000$, $\mu=0.5$, $\alpha=60^o$ and $\lambda=1.0$.}
\label{Fig10}
\end{figure}
\begin{figure}
\begin{tabular}{ll}
\hspace{-1.5in}\centerline{(a)\includegraphics[width=2.4in]{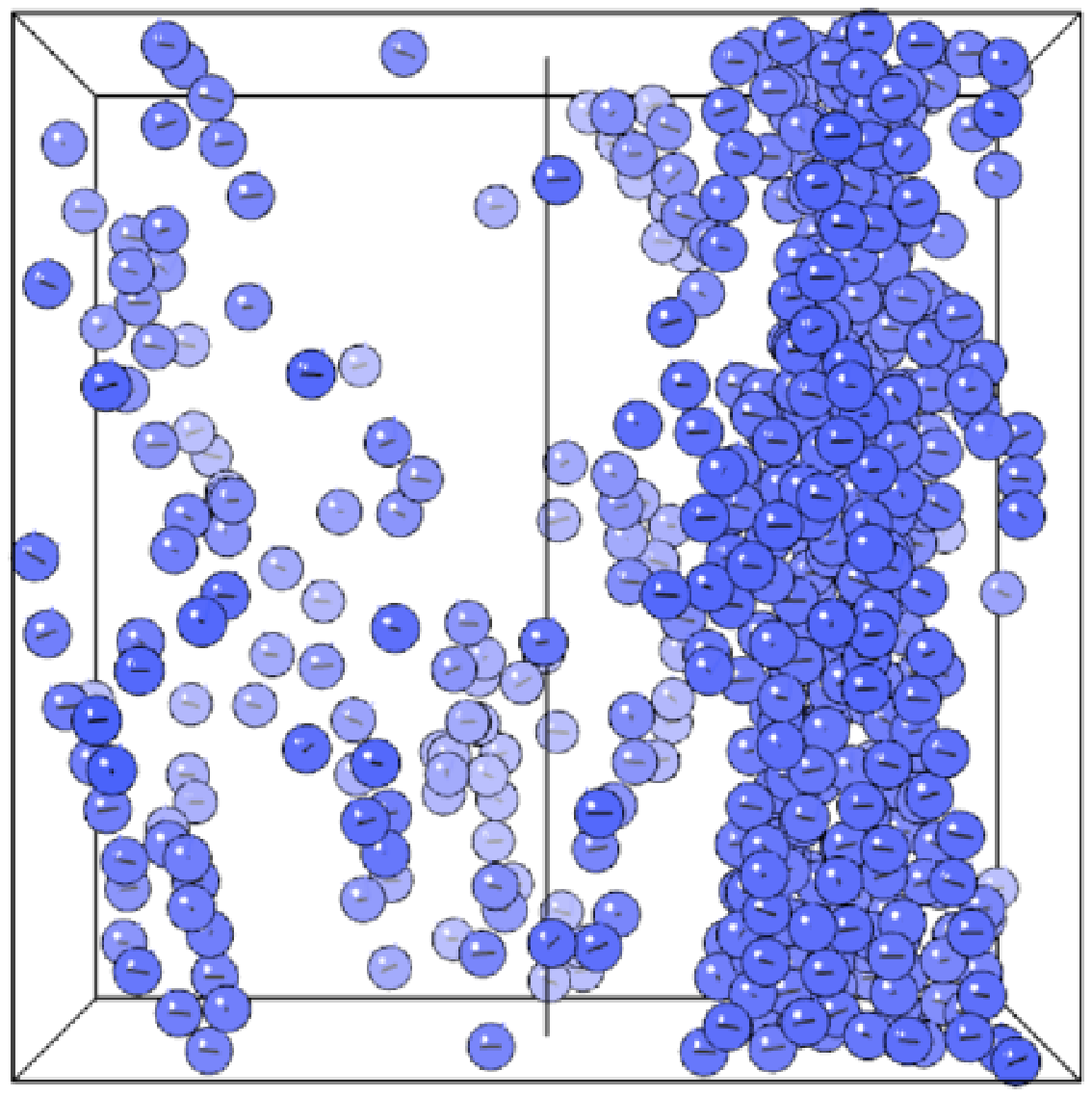}}&\hspace{-3.in}\centerline{(b)\includegraphics[width=2.4in]{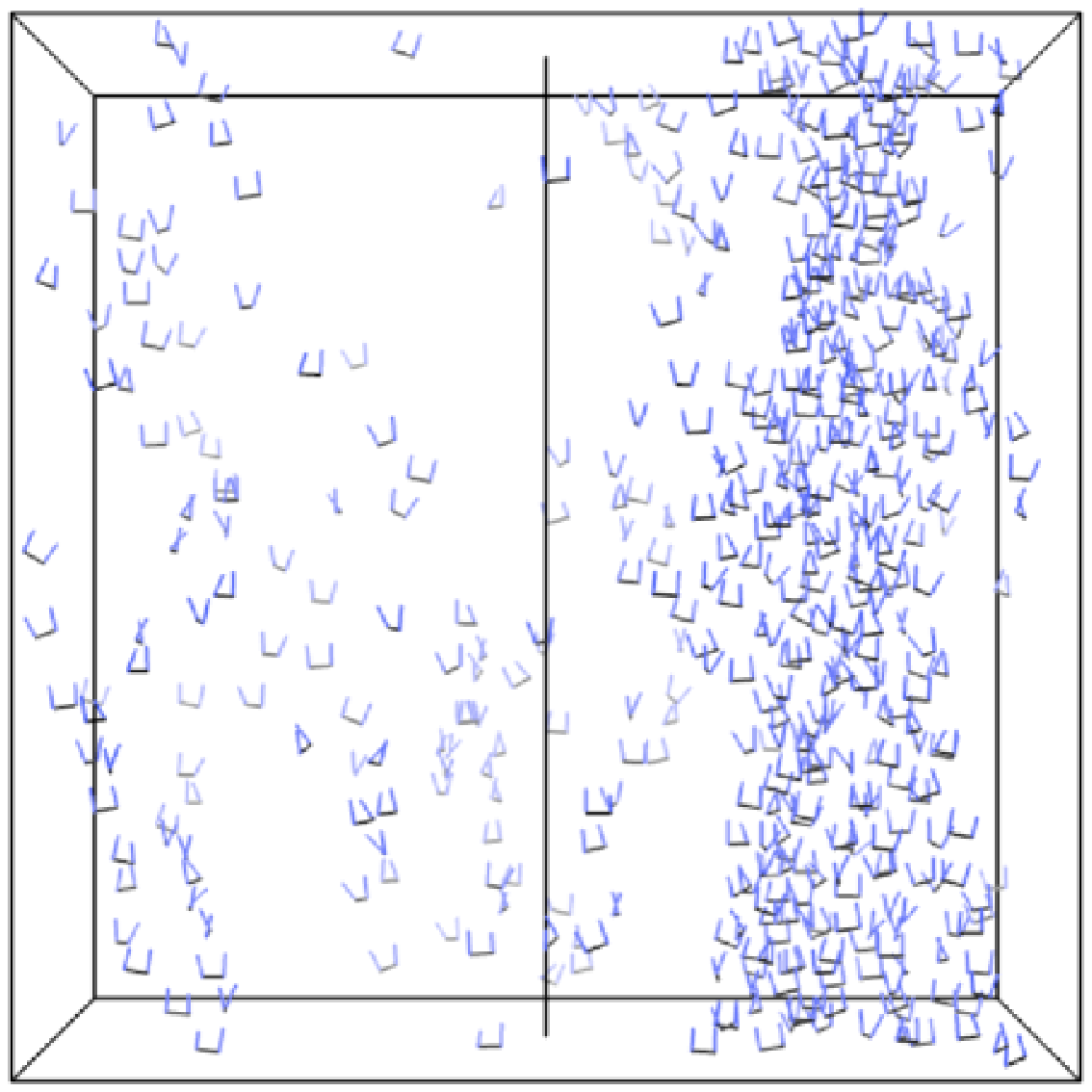}}\\
\hspace{-1.5in}\centerline{(c)\includegraphics[width=2.4in]{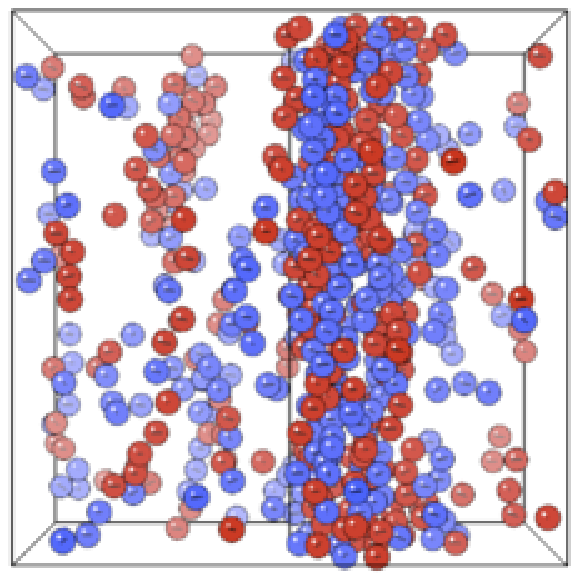}}&\hspace{-3.in}\centerline{(d)\includegraphics[width=2.4in]{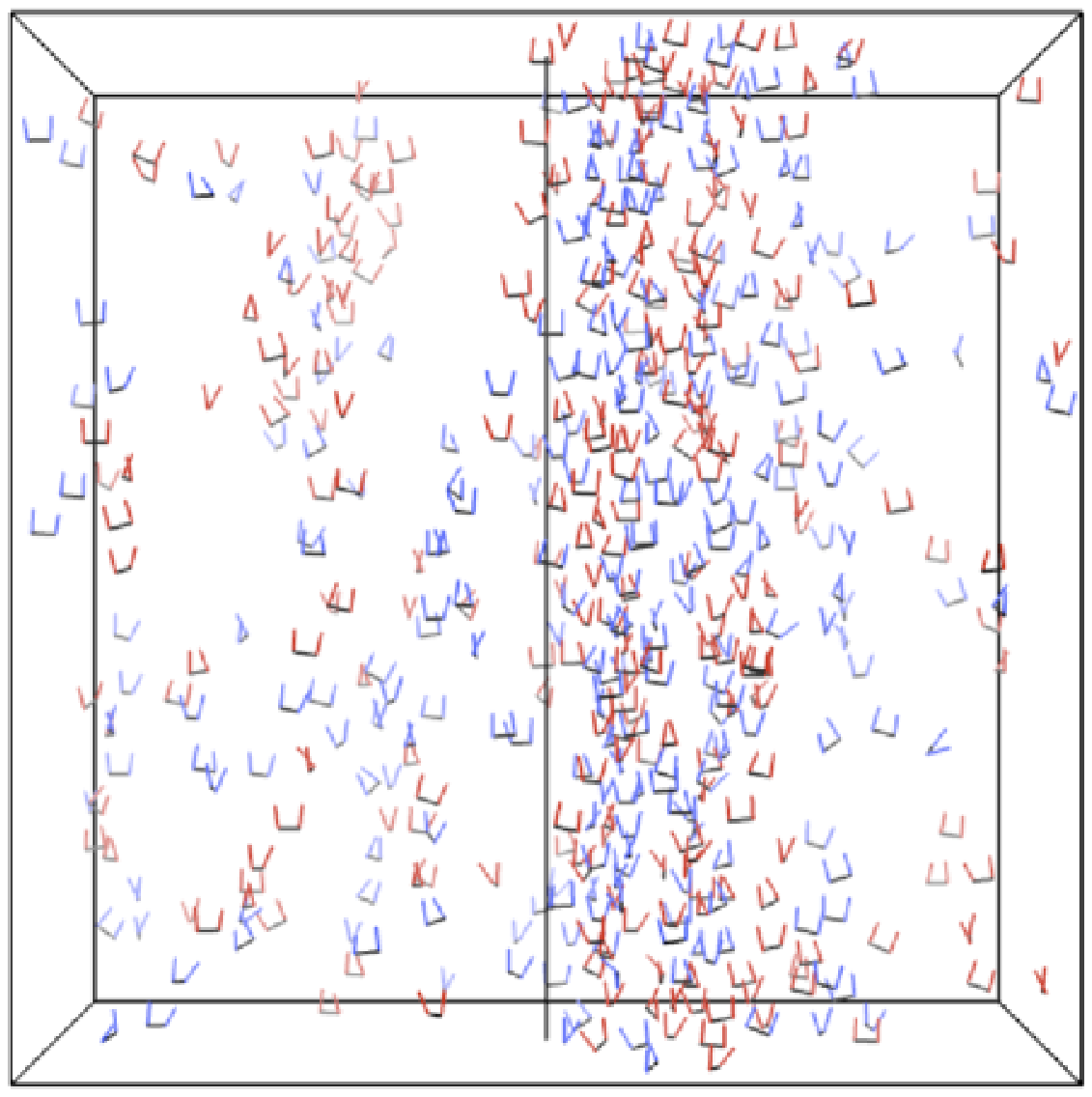}}\\
\hspace{-1.5in}\centerline{(e)\includegraphics[width=2.4in]{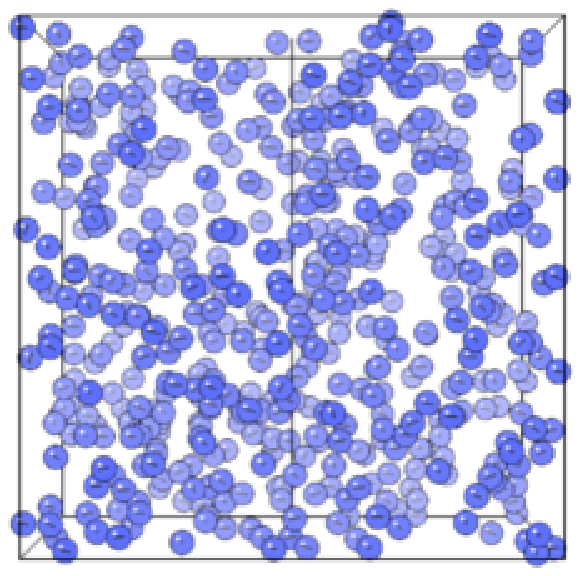}}&\hspace{-3.in}\centerline{(f)\includegraphics[width=2.4in]{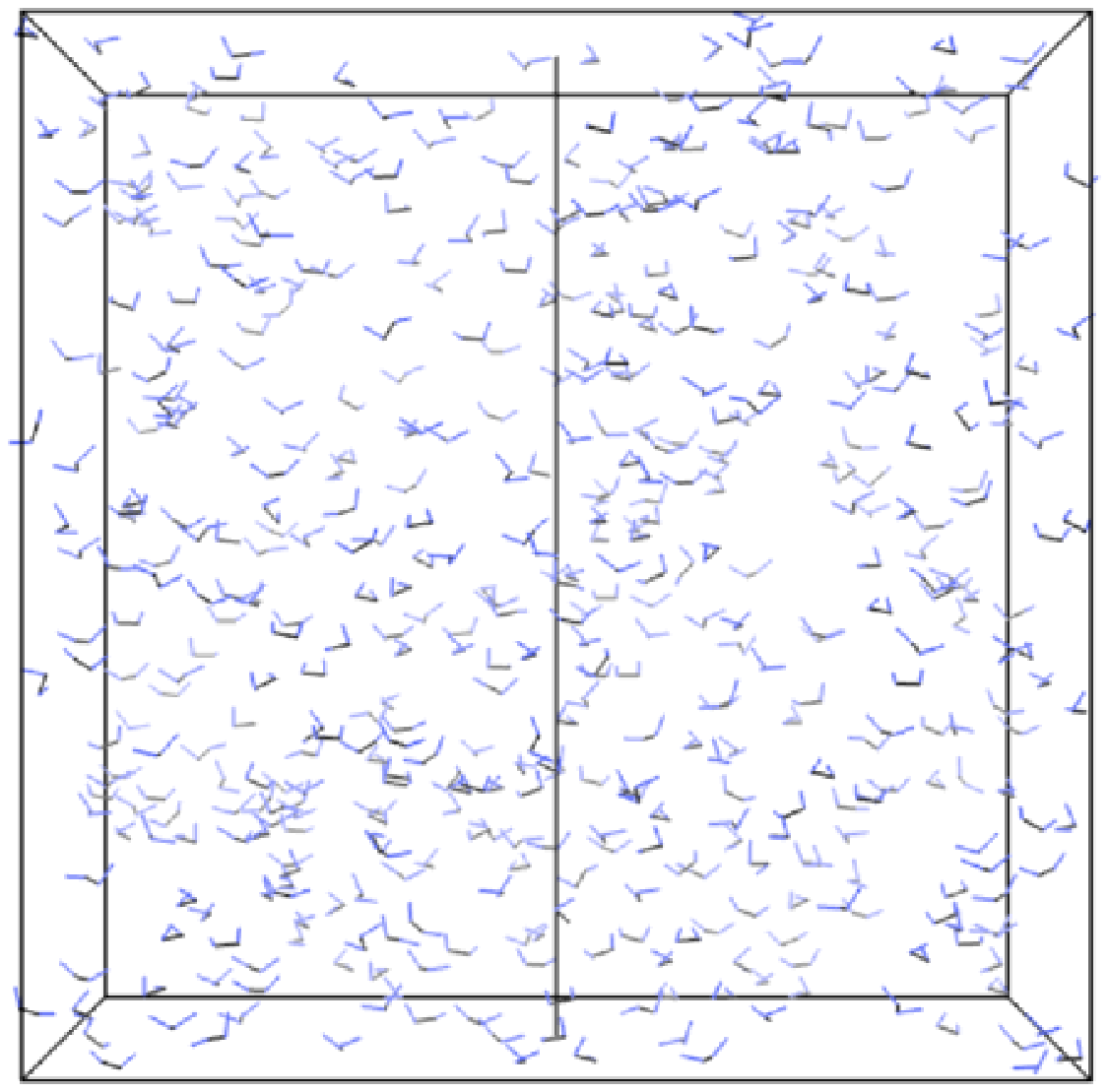}}
\end{tabular}
\caption{{\bf } Snapshots of homochiral-$\mbox{R}_\alpha$ and racemic systems in external electric field. For all systems  $N=512$, $\mu=1.0$, $\rho=0.05$, $\lambda=1.0$ and $ \bm{E} =E\hat{\bm{e}}_z$ ($E=10$) ; the solid vertical line indicates the direction of the field. (a-b): homochiral-$\mbox{R}_\alpha$  with $\alpha=35^o$ ; (c-d): racemic $\mbox{R}_\alpha$-$\mbox{S}_\alpha$ system with $\alpha=35^o$ and (e-f)homochiral-$\mbox{R}_\alpha$ with $\alpha=150^o$. R$_\alpha$-HS enantiomers are represented in blue and S$_\alpha$-HS enantiomers in red. In (b), (d) and (f), we represent, for the same configurations, molecular axis and segment in the direction of dipoles for each enantiomer ; dipoles of R$_\alpha$-HS enantiomers are represented in blue and dipoles of S$_\alpha$-HS enantiomers in red.}
\label{Fig11}
\end{figure}
\begin{equation}
\displaystyle U = \left< \sum_{a\neq b} E^{(a,b)} \right>
\label{UUU}
\end{equation}
where $E^{(a,b)}$ is the interaction energy between enantiomers $a$ and $b$ Eqs.(\ref{Tot_ab_int}-\ref{dist_iajb}) and $<.>$ is the average of the MC sampling. Some computation have been done with an external electric field $\bm{E}_0=E_0\bm{\hat{e}}_z$, the energy in the external field is computed as
\begin{table}
 \tbl{Preliminary results for Monte-Carlo computations in the isobaric $(NPT)$ ensemble of homochiral-$\mbox{R}_\alpha$ systems. These computations are performed with $N=1000$, $\mu=0.5$, $\alpha=60^o$ and $\lambda=1.0$. Notations are the same as in Table \ref{Table_MC_champ}, $p$ is the reduced pressure and $<\rho>$ is the average value of the density.}
{\begin{tabular}{@{}lccccc}\toprule
$p$ & $<\rho>$ & $\beta U/N$ & $S_+$ & $S_0$ & $S_-$\\
\colrule
&&&&&\\
0.1   & 0.086(2) & -0.08(1) & 0.02(1) & 0.00(1) & -0.02(1) \\
0.25 & 0.18(3)   & -0.17(1) & 0.03(1) & 0.00(1) & -0.03(1) \\
0.5   & 0.29(1)   & -0.27(2) & 0.02(1) & 0.00(1) & -0.02(1) \\
0.6   & 0.32(1)   & -0.30(2) & 0.02(1) & 0.00(1) & -0.02(1) \\
0.75 & 0.036(1) & -0.34(2) & 0.03(1) & 0.00(1) & -0.03(1) \\
0.8   & 0.37(1)   & -0.36(2) & 0.04(1) & 0.01(1) & -0.03(1) \\
1.0   & 0.41(1) & -0.39(2) & 0.03(1) & 0.00(1) & -0.03(1) \\
1.5   & 0.49(2) & -0.48(3) & 0.02(1) & 0.00(1) & -0.03(1) \\
2.0  & 0.55(3) & -0.54(3) & 0.02(1) & 0.01(1) & -0.03(1) \\
3.0  & 0.62(5) & -0.62(6) & 0.03(1) & 0.01(1) & -0.04(1) \\
4.0  & 0.66(6) & -0.7(1) & 0.03(1) & 0.01(1) & -0.03(1) \\
5.0  & 0.70(6) & -0.7(1) & 0.02(1) & 0.00(1) & -0.02(1) \\
\botrule
\end{tabular}}
\label{Table_MC_NPT}
\end{table}
\begin{figure}
\begin{tabular}{ll}
\hspace{-1.5in}\centerline{(a)\includegraphics[width=2.4in]{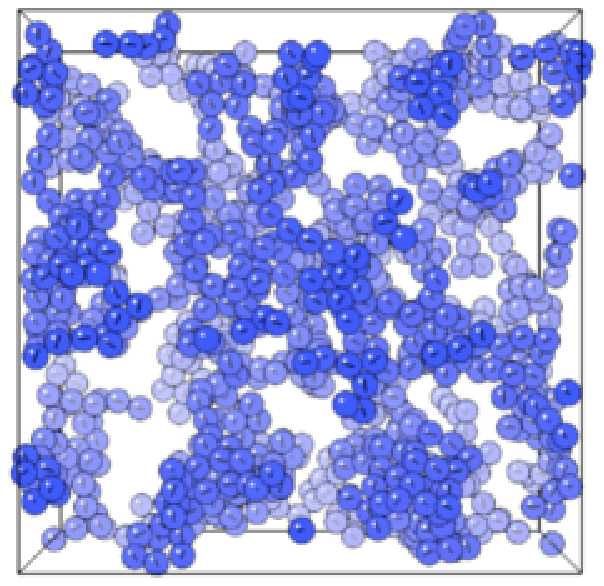}}&\hspace{-3.in}\centerline{(b)\includegraphics[width=2.4in]{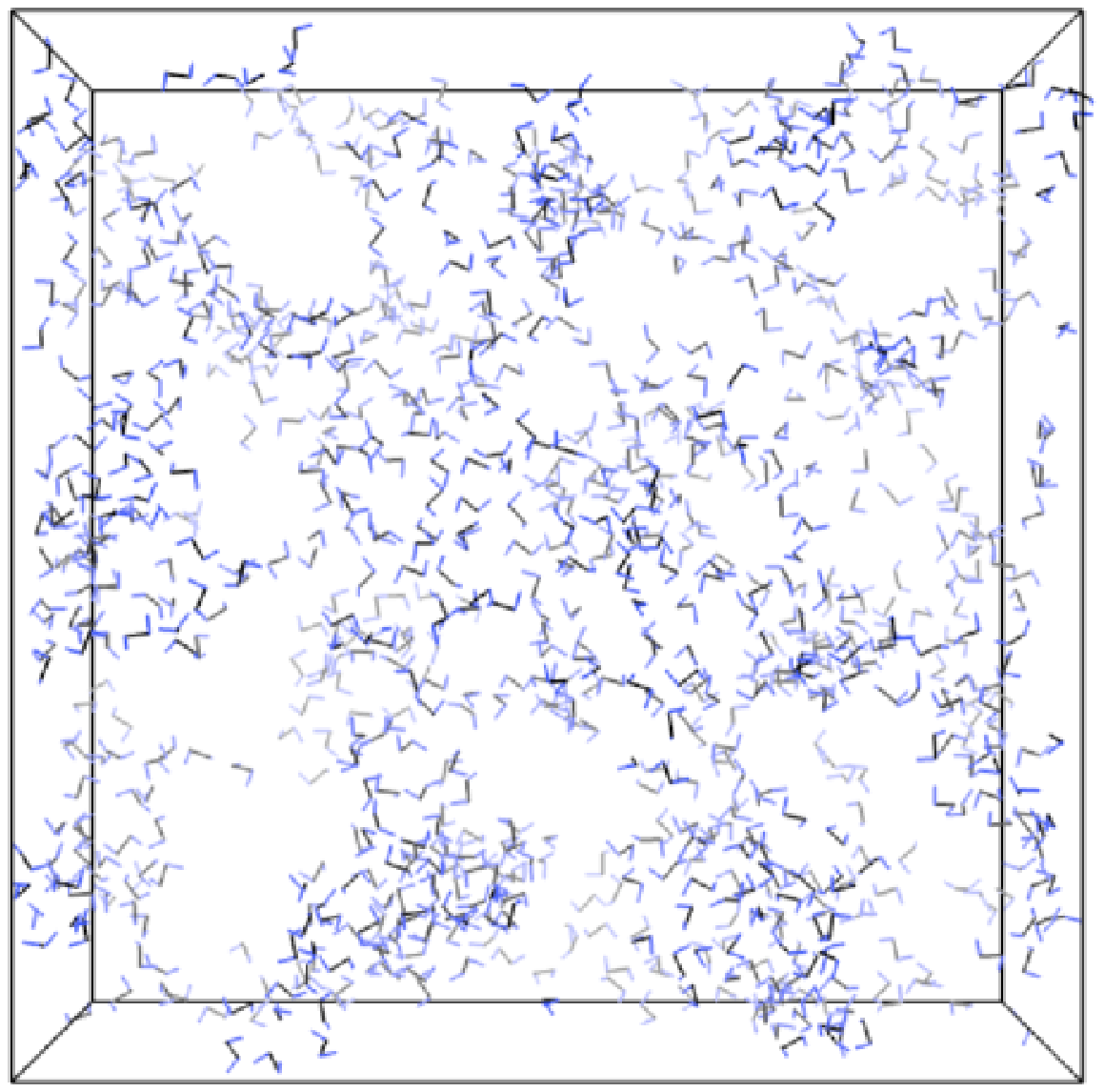}}\\
\hspace{-1.5in}\centerline{(c)\includegraphics[width=2.4in]{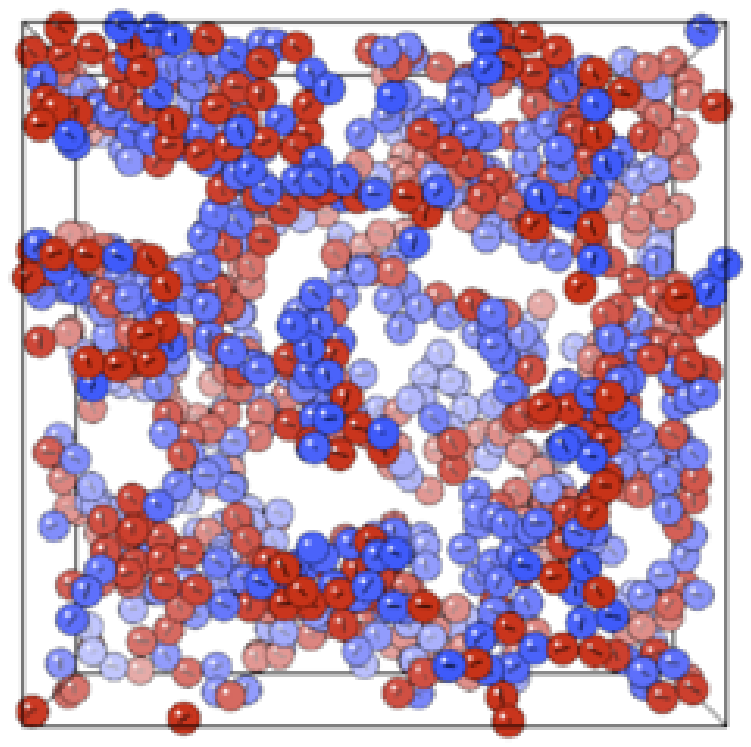}}&\hspace{-3.in}\centerline{(d)\includegraphics[width=2.4in]{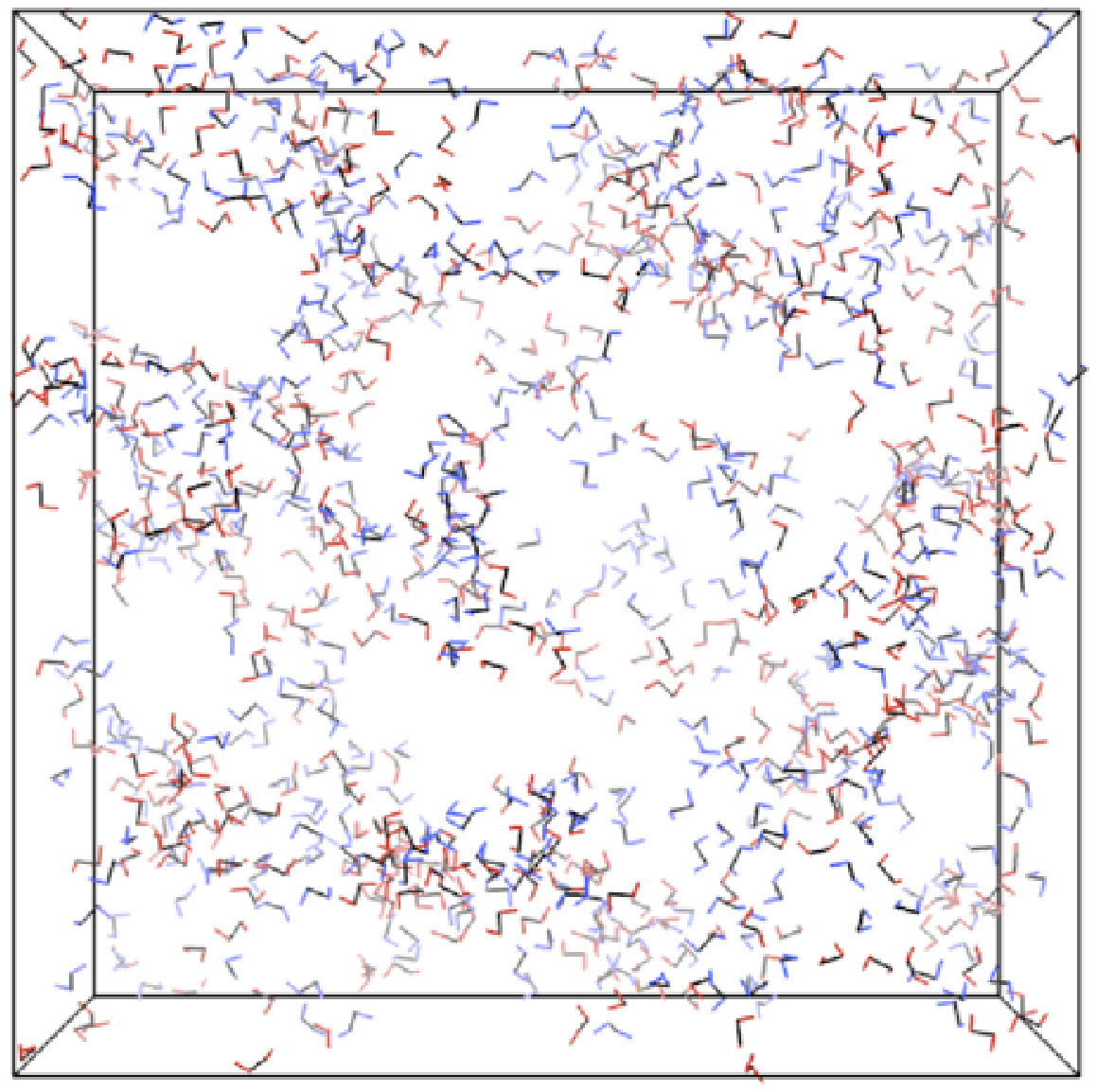}}
\end{tabular}
\caption{{\bf } Snapshots of chiral dipolar hard sphere systems for $N=1000$, $\mu=2.0$, $\rho=0.1$, $\alpha=120^o$ and $\lambda=1.0$. The notations are the same as those in Figs.\ref{Fig11}. (a-b): homochiral-$\mbox{R}_\alpha$ system ; (c-d): racemic $\mbox{R}_\alpha$-$\mbox{S}_\alpha$ system. In (a) and (c) hard spheres are represented as balls and molecular axis $O_1O_2$ by a black segment. No external electric field are applied to the systems.}
\label{Fig12}
\end{figure}
\begin{figure}
\centerline{\includegraphics[width=5.5in]{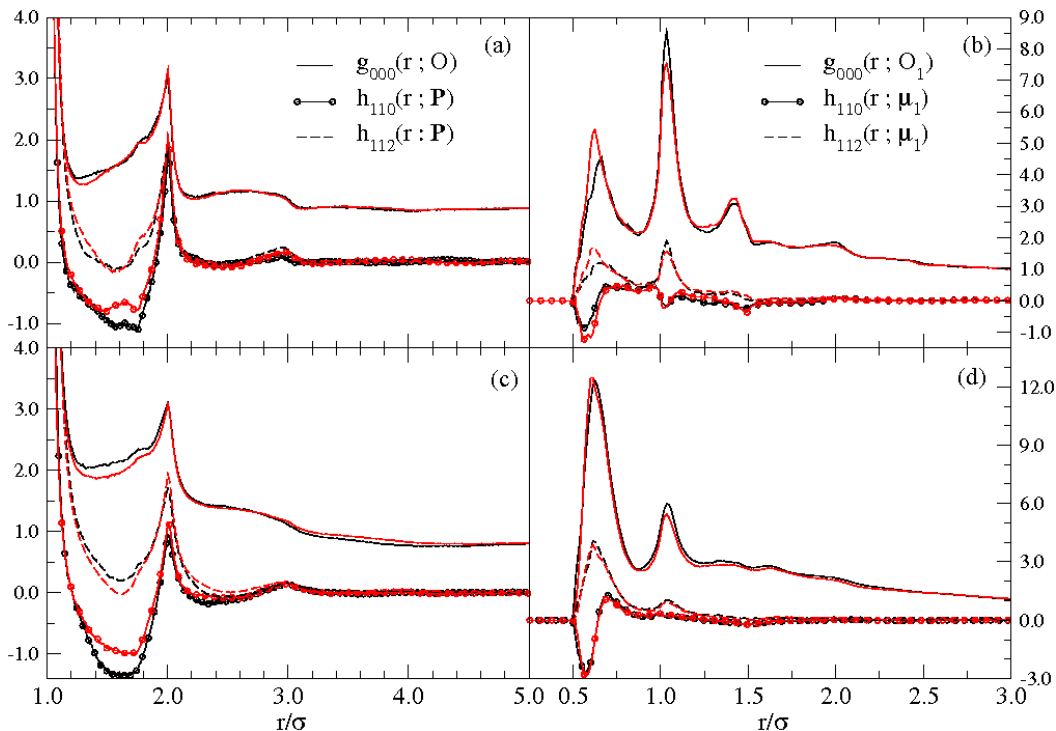}}
\caption{{\bf } Representation of projections of molecular pair distribution functions computed by Monte Carlo sampling of the $(NVT)$ ensemble of racemic $\mbox{R}_\alpha$-$\mbox{S}_\alpha$ systems. In these computation, $N=1000$, $\mu=1.5$, $\rho=0.1$ and $\alpha=30^o$ ; (a-b) $\lambda =1.0$ and (c-d) $\lambda=0.75$. The functions represented in (a) and (c) are computed for the set $(O,\bm{\hat{P}})$ and the functions represented in  (b) and (d) for the set $(O_1,\bm{\hat{\mu}}_1)$. In all figures, pair distributions functions for {\it Like-Like} pairs of molecules are represented in black and those for {\it Like-Unlike} pairs in red.}
\label{Fig13}
\end{figure}
\begin{equation}
\displaystyle V= \left< -\sum_{a=1}^{N}\bm{P}^{(a)}.\bm{E}_0\right> =  \left< -\sum_{a=1}^{N}(\bm{\mu}_1^{(a)}+\bm{\mu}_2^{(a)}).\bm{E}_0\right>
\label{UUU}
\end{equation}
Possible orientational order can be established by computing the eigenvalues of the tensor 
\begin{equation}
\displaystyle Q_{\alpha \beta} = \frac{1}{2N} \sum_{i=1}^{N} (3 \hat{x}_{\alpha}^i \hat{x}_{\beta}^i-\delta_{\alpha \beta})
\label{OrderPara_Nen}
\end{equation}
where $\hat{x}_{\alpha}^i$ is the cartesian component of a unit vector $\bm{\hat{x}}$ of molecule $i$. The largest eigenvalue $S_+$ of $Q_{\alpha \beta}$ is used as an order parameter for the isotropic-nematic phase transition in liquid crystals. It is also worthwhile to note that if, all unit vectors $\bm{\hat{x}}$ are perpendicular to a given fixed direction then, the eigenvalues are $S_+=S_0=1/4$ and $S_-=-1/2$. In the present work, we have restricted the computation of $Q_{\alpha \beta}$ for the molecular axis $\bm{\hat{u}}$ of enantiomers.\\
To study the structure of the systems we have computed the angular projections on rotational invariants of the molecular pair distribution function \cite{Wertheim:71,Blum:72, Lomba:00,Weis:02a} ; in the present work, we have computed only $g_{000}(r)$, $h_{110}(r)$ and $h_{112}(r)$ for two sets of points and vectors. The first set is the molecule center and the orientation of the total dipole moment : $(O,\bm{\hat{P}})$ ; the second set is the site $O_1$ and the orientation of the dipole $\bm{\mu}_1$ : $(O_1,\bm{\hat{\mu}}_1)$. To study the influence of the chirodiastaltic interaction on structure of mixtures of both enantiomers, it will be necessary to compute projections of the pair distribution functions on rotational invariants of higher order since the discriminatory interaction involves dipole-quadrupole interactions.\\
Table \ref{Table_MC_champ}, summarizes average energies and eigenvalues of $Q_{\alpha \beta}$ Eq.(\ref{OrderPara_Nen}) for homochiral-$\mbox{R}_\alpha$ and racemic systems in external electric field $\bm{E} =E\hat{\bm{z}}$ ($E=10$), for several values of $\alpha$ and $\lambda$. On Table \ref{Table_MC_NPT}, we report some preliminary results of Monte-Carlo computations in the isobaric $(NPT)$ ensemble of homochiral-$\mbox{R}_\alpha$ systems and on Fig.\ref{Fig10}, we show $g_{000}(r)$ obtained with these computations.\\
\begin{figure}
\begin{tabular}{ll}
\hspace{-1.5in}\centerline{(a)\includegraphics[width=2.4in]{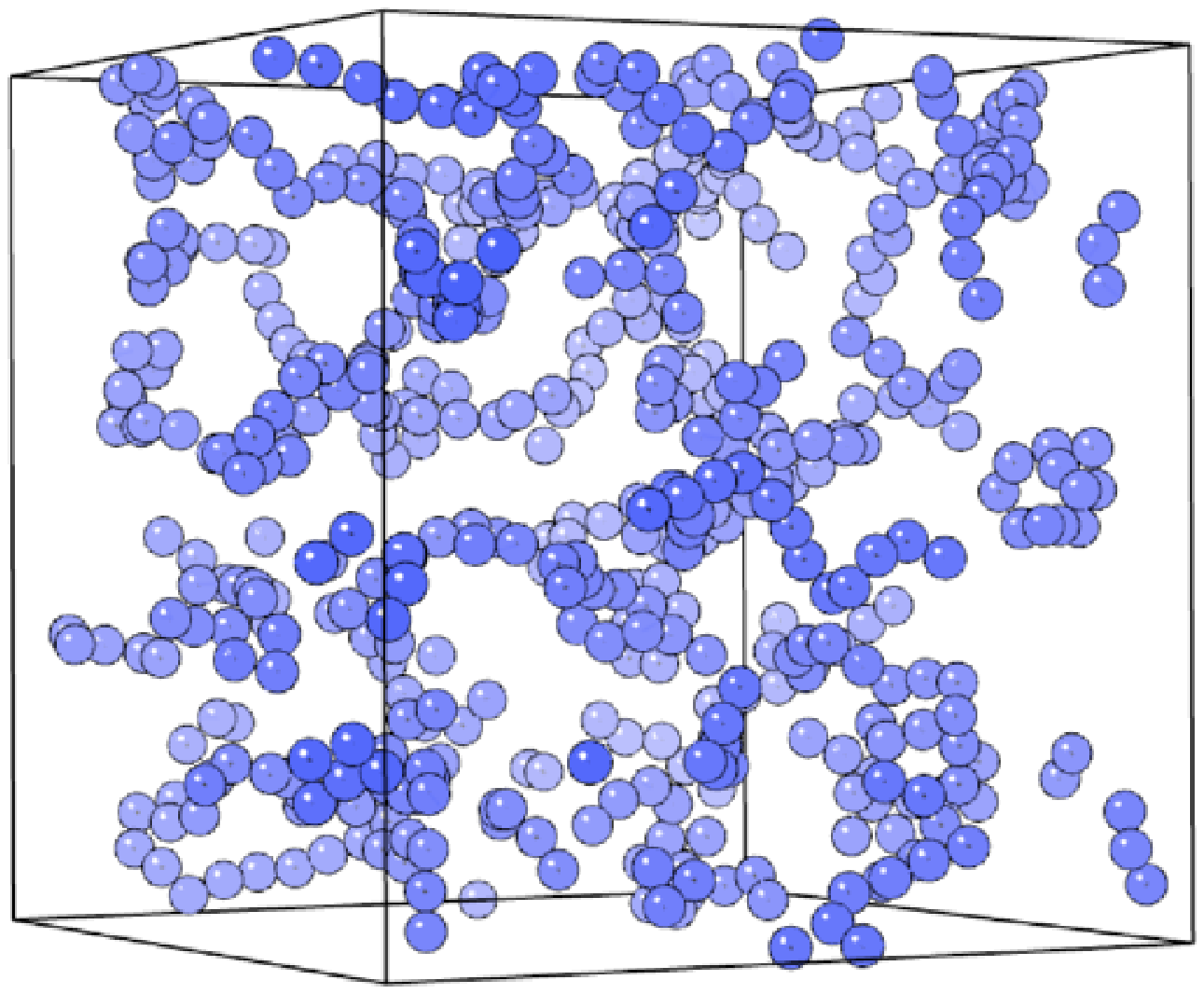}}&\hspace{-3.in}\centerline{(b)\includegraphics[width=2.4in]{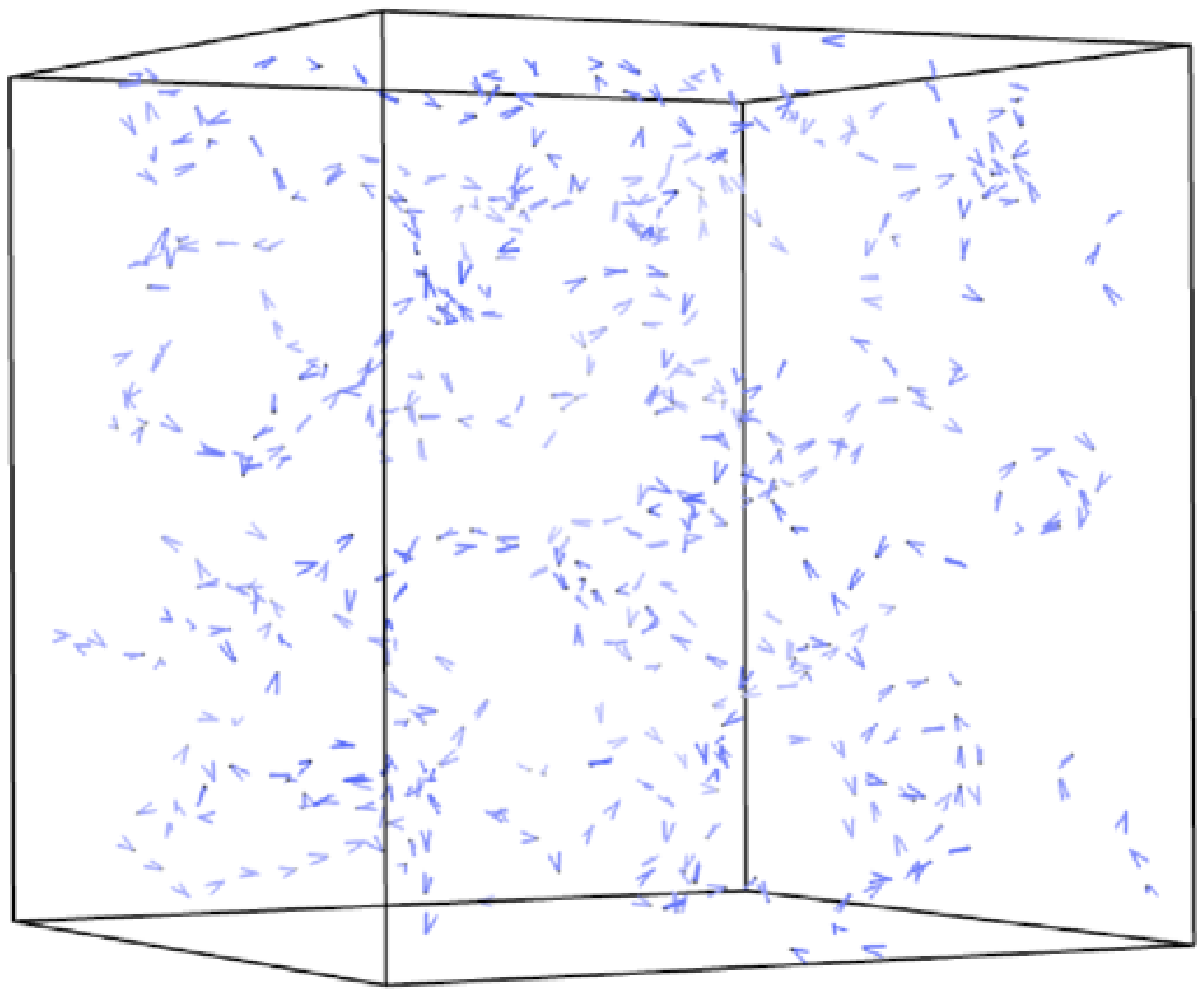}}\\
\hspace{-1.5in}\centerline{(c)\includegraphics[width=2.4in]{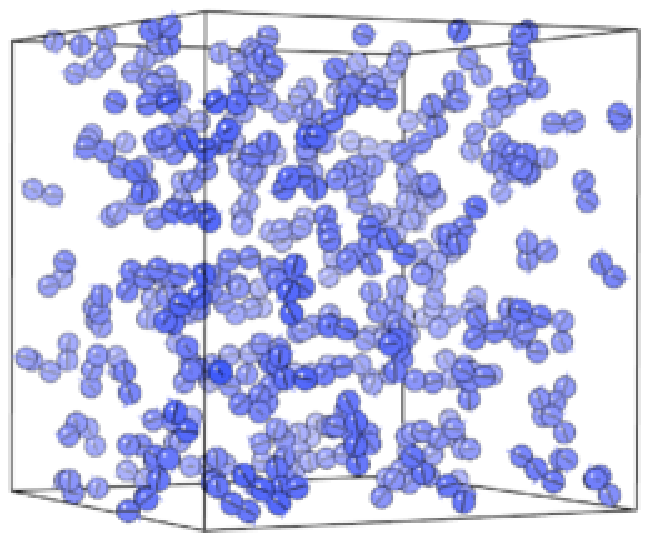}}&\hspace{-3.in}\centerline{(d)\includegraphics[width=2.4in]{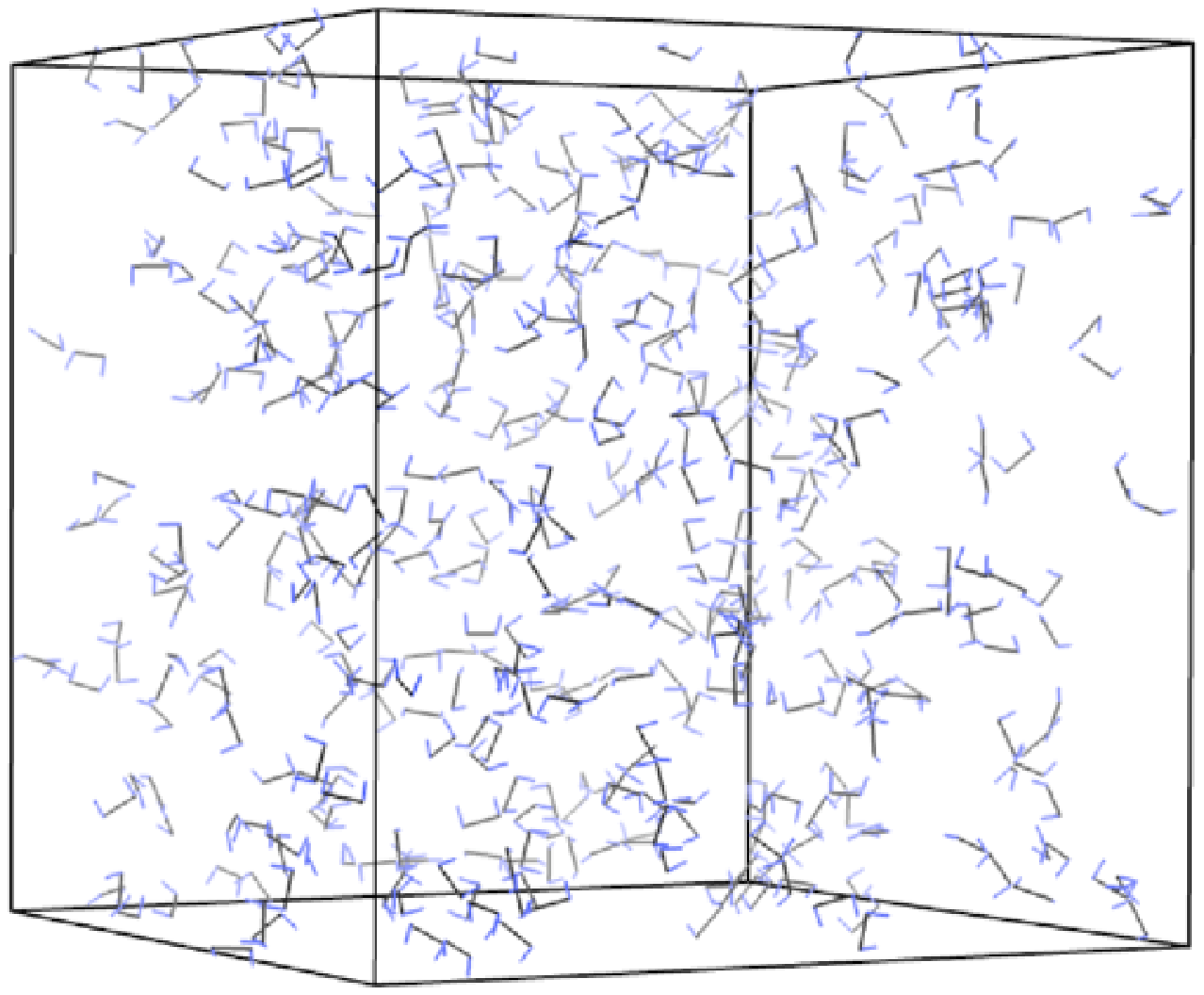}}
\end{tabular}
\caption{{\bf } Snapshots of chiral dipolar hard sphere systems for homochiral-$\mbox{R}_\alpha$ system with $N=512$, $\mu=1.5$, $\rho=0.05$, $\alpha=30^o$ and $\lambda=1.0$ and for two value of $L$. The notations are the same as those in Figs.\ref{Fig11}. (a-b): $2L=0.05\sigma$ ; (c-d): $2L=0.95\sigma$. In (a) and (c) hard spheres are represented as balls and molecular axis $O_1O_2$ by a black segment. No external electric field are applied to the systems.}
\label{Fig14}
\end{figure}
On Fig.\ref{Fig11}, we give few snapshots of homochiral-$\mbox{R}_\alpha$ and racemic systems in an external field. The results reported on table \ref{Table_MC_champ} show that energies and the nematic order parameters for the molecular axis are almost independent on the composition of systems. The eigenvalues $S_+$, $S_0$ and $S_-$ show that on average the molecular axis of enantiomers are perpendicular to the direction of the electric field, this results is quite obvious since the total dipole moment, as defined in Eqs.(\ref{frame},\ref{dipole}), is perpendicular to the molecular axis this tendency decreases as $\alpha$ increases, since the total dipole moment decreases. As shown on the snapshots in Fig.(\ref{Fig11}), if $\alpha\lesssim 60^o$ molecules tend to condense in a large columnar cluster, for both homochiral-$\mbox{R}_\alpha$ and racemic systems in an external electric field (cf. Fig.\ref{Fig11} (a-d)) ; if $\alpha\gtrsim 70^o$ the columnar cluster does not form (cf. Fig.\ref{Fig11} (e-f)). There are at least two facts that may help to understand qualitatively the disappearance of the columnar clusters as $\alpha$ increases, while $\lambda$ and $\mu$ are fixed. First, as $\alpha$ increases, the total dipole of each molecule decreases and thus formations of chains in the direction of the external field is less favoured. Second, for large enough $\alpha$, configurations with two dipoles belonging to two different molecules may be more easily found, on average, in an antiparallel configuration in a plane perpendicular to the field, while the total dipole of both molecules can still be parallel to the field. At present, it is not clear how the columnar clusters form and which mecanism favour their formation ; further investigations are needed to be conclusive. Homochiral-$\mbox{R}_\alpha$ and racemic $\mbox{R}_\alpha$-$\mbox{S}_\alpha$ systems in an external electric field exhibit the same properties, however, a close inspection of the snapshots of Fig.\ref{Fig11} (a-d) shows that the internal structure of the columnar clusters is not the same in homochiral and racemic systems. In homochiral systems, the molecules in the columnar cluster have an helical arrangement (in agreement with the results shown on Fig.\ref{Fig6}) ; in racemic systems, these helical arrangement in the columnar clusters are less marked because of the {\it Like-Unlike\/} interactions. To obtain more quantitative descriptions of the internal structure of the columnar clusters, some particular order parameters and distribution functions have to be defined.\\
If there is no external electric field, columnar clusters are not formed instead some complicated structures are formed at low density and large enough $\mu$ ; this has occured for all values of $\alpha$ that have been considered in these prelyminary results. On Fig.\ref{Fig12}, snapshots of homochiral-$\mbox{R}_\alpha$ and racemic $\mbox{R}_\alpha$-$\mbox{S}_\alpha$ systems are represented, for these systems $\alpha=120^o$ and no external electric fileds are applied. Again, in these computations homochiral and racemic systems behave very similarly, however, an inspection of the pair distribution functions show that there is small differences between pair distributions functions for {\it Like-Like} pairs of molecules and those for {\it Like-Unlike} pairs (see for instance Fig.\ref{Fig13}).
The comparison between pair distribution functions computed for $\lambda=1.0$ and $0.75$ shows that this parameter have an important influence on the local structure (cf. Fig.\ref{Fig13}). In particular, when $\lambda\neq 1.0$, interactions between the dipoles $\bm{\mu}_1$ of two enantiomers are favoured and so an asymmetric arrangement of the molecular axis.\\
There are three important parameters in this chiral model : $\lambda$, $\alpha$ and $L$. So far in this section in the simulations reported (and also in numerical computations done in subsection 2.2),  we have set always $L/\sigma=1/4$ and showed how energy minima and structures depend on  $\lambda$ and $\alpha$. They will depend also on the ratio $L/\sigma$ : for instance, if $L/\sigma$ is small, chains with colinear aligned total dipole moment will be favoured, while, if $L/\sigma$ is close to 1, structures with colinear aligned molecular axis will be favoured, this behaviour is illustrated on Fig.\ref{Fig14}.

\section{\label{sec:level4} Perspectives}

A full determination of the structures and clusters at low densities and temperatures will require a larger sampling of the phase space ; this can be achieved with longer runs and by implementing trial cluster moves \cite{Frenkel:04,Whitelam:07,Almarza:07}. Nematic and helical-cholesteric order parameter and also the bond orientational order parameters \cite{Nelson:81} will be useful to investigate the properties of structure at low density as function of the {\it enantiomeric excess\/} and model parameters, homochiral systems have interest by themselves. Same order parameters and simulations in the isobaric ensemble $(NPT)$ will serve to study the crystal phases of homochiral systems and also as functions of $ee$. Works and computations in these directions are already in progress.\\
Another chiral model may be defined from the computation of the multipole moments done in section 2, and in agreement with earlier studies on chiral discrimination due to multipole-multipole interactions \cite{Craig:76,Craig:74,Craig:75}. According to Eqs. (4,7), the three components, $Q_{10}$, $Q_{21}$ and $Q_{22}$ are independent, thus an hard sphere model with dipole-dipole, dipole-quadrupole and quadrupole-quadrupole interactions \cite{Aguado:03,Laino:08} can be built by using these three multipole components as parameter for the model instead of $(\lambda,\alpha,L)$. Such model will be slightly different from the bi-dipolar hard sphere model (and also from the four charges model), since higher order multipole-multipole interactions will not be taken into account in this model. Such systems has not been studied in the present work.\\
With the bidipolar model, one may easily build a model of chiral liquid crystal by changing the short ranged steric repulsion by replacing the hard sphere potential by a spherocylinder \cite{Vieillard:72,Frenkel:85} or ellipsoid hard potential \cite{Vieillard:74,Veerman:90}, or even by a soft Gay-Berne potential \cite{Gay:81,deMiguel:92}. For these chiral liquid crystal models with two dipoles, all the results of section 2 hold and in particular : the computations of the multipole components and interaction energies between to enantiomers. The modifications of the phase diagram of these systems with the introduction of these chirodiastaltic interactions can be of particular interest for the physics of liquid crystals.\\    
 
\section*{Acknowledgement}
The author acknowledges computation facilities provided by the {\it Institut du D\'eveloppement et des Ressources en Informatique Scientifique\/} (IDRIS) under projects 0682104 and 0992104.\\
In the middle of nineties Jean-Jacques was my PhD supervisor  ; I have to say that I have been very lucky to benefit of his broad knowledge in the physics of liquids and in computer simulations. Therefore, I am extremely glad to dedicate this paper to Jean-Jacques Weis in this special issue of {\it Molecular physics\/} in his honour.

\newpage
\appendices
\section{Multipole expansion for a dipole distribution.}
In this appendix, we give general formulas for the multipole expansion of a dipole distribution $\bm{\mu}(\bm{r})$ located in a finite region $\Omega$ of the space. From classical electromagnetism theory \cite{Jackson:75,Gray:84} and choosing the origin inside $\Omega$,  the electrostatic potential $\Phi(\bm{r})$, due to the dipole distribution in $\Omega$, is given by
\begin{equation}
\begin{array}{ll}
\displaystyle \Phi(\bm{r})&\displaystyle=\int_{\Omega}\frac{\bm{\mu}(\bm{r}').(\bm{r}-\bm{r}')}{\mid \bm{r}-\bm{r}'\mid^3}d\bm{r}'=-\int_{\Omega}\bm{\mu}(\bm{r}').\bm{\nabla}_{\bm{r}}\left(\frac{1}{\mid \bm{r}-\bm{r}'\mid}\right)d\bm{r}'\\
&\\
&\displaystyle = \int_{\Omega}\bm{\mu}(\bm{r}').\bm{\nabla}_{\bm{r}'}\left(\frac{1}{\mid \bm{r}-\bm{r}'\mid}\right)d\bm{r}'
\end{array}
\label{EqA1}
\end{equation}
Since we are interested in the potential at $\bm{r}$ outside the dipole distribution, we set $r>r'$ and $\hat{\bm{r}}'$ and $\hat{\bm{r}}$, the unitary vectors that define directions of $\bm{r}'$ and $\bm{r}$. The relation 
\begin{equation}
\displaystyle \frac{1}{\mid \bm{r}-\bm{r}'\mid} =\sum_{l=0}^{\infty}\sum_{m=-l}^{m=l}\left(\frac{4\pi}{2l+1}\right)\frac{r'^l}{r^{(l+1)}}\mbox{Y}_{lm}(\hat{\bm{r}}')\mbox{Y}^{*}_{lm}(\hat{\bm{r}})
\label{EqA2}
\end{equation}
is reported in Eq.(\ref{EqA1}) and the multipole expansion for a dipole distribution is given by
\begin{equation}
\displaystyle \Phi(\bm{r})=\sum_{l=0}^{\infty}\sum_{m=-l}^{m=l}\left(\frac{4\pi}{2l+1}\right)Q_{lm}\frac{\mbox{Y}^{*}_{lm}(\hat{\bm{r}})}{r^{(l+1)}}
\label{EqA3}
\end{equation}
with $Q_{lm}$ the $m^{th}$ component of the spherical multipole moment tensor of order $l$ defined by
\begin{equation}
\displaystyle Q_{lm}=\int_{\Omega}\bm{\mu}(\bm{r}').\bm{\nabla}_{\bm{r}'}\left[r'^l \mbox{ Y}_{lm}(\hat{\bm{r}}')\right]d\bm{r}'
\label{EqA4}
\end{equation}
Eqs. (\ref{EqA3}-\ref{EqA4}) are the multipole expansion for a dipole distribution. In the following, we give explicitly the $Q_{lm}$ up to $l=3$ for any dipole distribution. From Eq.(\ref{EqA4}), and since any dipole distribution can also be represented as a real charge density, it is easy to show that
\begin{equation}
\displaystyle Q_{l,-m} = (-1)^m Q^*_{l,m}
\label{EqA5}
\end{equation}
\subsection{The total charge in $\Omega$ : $Q_{00}$.}
From Eq.(\ref{EqA4}), we have
\begin{equation}
\displaystyle Q_{00}=\int_{\Omega}\bm{\mu}(\bm{r}).\bm{\nabla}_{\bm{r}}\left[\mbox{ Y}_{00}(\hat{\bm{r}})\right]d\bm{r}=0
\label{EqA6}
\end{equation}
Of course, this result is obvious for any dipole distribution.
\subsection{The dipole moment tensor : $Q_{1m}$.}
The components $Q_{11}$ and $Q_{10}$ are given by 
\begin{equation}
\left\{ \begin{array}{ll}
\displaystyle Q_{11} &\displaystyle=-\sqrt{\frac{3}{8\pi}}\left( \int_{\Omega}\bm{\mu}(\bm{r}).\bm{\nabla}_{\bm{r}}\left[r \sin\theta e^{i\phi}\right] d\bm{r}\right) \\
&\\
\displaystyle Q_{10} &\displaystyle=\sqrt{\frac{3}{4\pi}}\left( \int_{\Omega}\bm{\mu}(\bm{r}).\bm{\nabla}_{\bm{r}}\left[r \cos\theta \right] d\bm{r}\right)
\end{array}
\right.
\label{EqA7}
\end{equation}
thus, we find
\begin{equation}
\left\{ \begin{array}{ll}
\displaystyle Q_{11} &\displaystyle=-\sqrt{\frac{3}{8\pi}}\left[ P_x + i P_y \right]\\
&\\
\displaystyle Q_{10} &\displaystyle=\sqrt{\frac{3}{4\pi}}P_z
\end{array}
\right.
\label{EqA8}
\end{equation}
Where $\bm{P}$ is the total dipole in $\Omega$ given by
\begin{center}
$\displaystyle \bm{P}=\int_{\Omega}\bm{\mu}(\bm{r})d\bm{r}$ 
\end{center}
This result is exactly the same as the one obtained for a charge distribution $\rho(\bm{r})$ with the total dipole defined by 
\begin{center}
$\displaystyle \bm{P}=\int_{\Omega}\bm{r}\rho(\bm{r})d\bm{r}$ 
\end{center}
\subsection{The quadrupole moment : $Q_{2m}$.}
The same derivation gives for the spherical components of the quadrupole
\begin{equation}
\left\{ \begin{array}{ll}
\displaystyle Q_{22} &\displaystyle=\frac{1}{2}\sqrt{\frac{15}{2\pi}}\left[\int_\Omega x\mu_x d\bm{r}-\int_\Omega y\mu_y d\bm{r}+i\int_\Omega (y\mu_x+x\mu_y) d\bm{r}\right]\\
&\\
\displaystyle Q_{21} &\displaystyle= -\sqrt{\frac{15}{8\pi}}\left[\int_\Omega (x\mu_z+z\mu_x) d\bm{r} +i\int_\Omega (y\mu_z+z\mu_y) d\bm{r} \right]\\
&\\
\displaystyle Q_{20} &\displaystyle= \sqrt{\frac{5}{4\pi}}\left[2\int_\Omega z\mu_z d\bm{r}-\int_\Omega x\mu_x d\bm{r}-\int_\Omega y\mu_y d\bm{r} \right]
\end{array}
\right.
\label{EqA10}
\end{equation}
The relations between spherical and cartesian components are given by\cite{Gray:84},
\begin{equation}
\left\{ \begin{array}{ll}
\displaystyle Q_{22}&\displaystyle= \sqrt{\frac{5}{24\pi}}\left(q_{xx}-q_{yy}+2iq_{xy}\right)\\
&\\
\displaystyle Q_{21} &\displaystyle= -\sqrt{\frac{5}{6\pi}}\left(q_{xz}+iq_{yz}\right)\\
&\\
\displaystyle Q_{20} &\displaystyle=\sqrt{\frac{5}{4\pi}}q_{zz}
\end{array}
\right.
\label{EqA11}
\end{equation}
and the traceless quadrupole moment tensor in cartesian coordinates can be written as
\begin{equation}
\displaystyle q_{\alpha\beta}=\int_\Omega\left(\frac{3}{2}(r_\alpha\mu_\beta+r_\beta\mu_\alpha)-(\bm{r}.\mu)\delta_{\alpha\beta}\right)d\bm{r}
\label{EqA12}
\end{equation}
\subsection{The octopole moment : $Q_{3m}$.}
The spherical components of the octopole moment may also be computed from Eq.(\ref{EqA4}) ; after some algebra, we found
\begin{equation}
\left\{ \begin{array}{ll}
\displaystyle Q_{33} &\displaystyle= -\frac{3}{4}\sqrt{\frac{35}{4\pi}} \int_\Omega (\mu_x+i\mu_y)(x+iy)^2 d\bm{r}  \\
&\\
\displaystyle Q_{32} &\displaystyle=\frac{1}{2}\sqrt{\frac{105}{8\pi}}\left[ \int_\Omega \mu_z (x+iy)^2 d\bm{r} +2 \int_\Omega (\mu_x+i\mu_y)z (x+iy)  d\bm{r} \right]\\
&\\
\displaystyle Q_{31} &\displaystyle=-\frac{1}{8}\sqrt{\frac{21}{4\pi}}\left[8 \int_\Omega\mu_z z (x+iy)d\bm{r}+ \int_\Omega (\mu_x-i\mu_y) (x-iy)^2  d\bm{r} \right.\\
&\\
&\displaystyle\left.\mbox{                    } + 4\int_\Omega(\mu_x(z^2-x^2)+i\mu_y(z^2-y^2))  d\bm{r}\right] \\
&\\
\displaystyle Q_{30} &\displaystyle=\frac{3}{2}\sqrt{\frac{7}{4\pi}}\left[ \int_\Omega \mu_z(2z^2-x^2-y^2) d\bm{r} - 2\int_\Omega z(\mu_x x+\mu_y y) d\bm{r}\right]
\end{array}
\right.
\label{EqA13}
\end{equation}
and the cartesian components of the octopole moment are given by
\begin{equation}
\begin{array}{ll}
\displaystyle \Omega_{\alpha\beta\gamma}=\frac{1}{2}\int_\Omega d\bm{r}\left[ \right.&\displaystyle\frac{5}{3}(\mu_\alpha r_\beta r_\gamma+r_\alpha\mu_\beta r_\gamma+r_\alpha r_\beta\mu_\gamma) \\
&\\
&\displaystyle - r^2(\mu_\alpha\delta_{\beta\gamma}+\mu_\beta\delta_{\alpha\gamma}+\mu_\gamma\delta_{\beta\alpha}) \\
&\\
&\displaystyle \left. -2(\bm{r}.\bm{\mu})(r_\alpha\delta_{\beta\gamma}+r_\beta\delta_{\alpha\gamma}+r_\gamma\delta_{\beta\alpha}) \right]

\end{array}
\label{EqA14}
\end{equation}

\label{lastpage}


\begin{thebibliography}{19}

\bibitem{Pasteur:1847} L. Pasteur, "{\it \'Etudes des Ph\'enom\`enes Relatifs \`a la Polarisation Rotatoires des Liquides\/}", thesis in chemistry and physics, Faculty of Sciences of Paris (August 23, 1847) ; L. Pasteur "{\it Recherches sur le Dimorphisme\/}", Annales de chimie et de physique, 3 s\'er., {\bf XXIII}, p.267-294 (1848) ; these two references are reprinted in {\it \OE uvres de Pasteur -Tome Premier : Dissym\'etrie Mol\'eculaire\/}, edited by Pasteur Vallery-Radot (\'Editions Masson, Paris, 1922).
\bibitem{Cahn:56} R.-S. Cahn, C.-K. Ingold and V. Prelog, {\it Experientia}, {\bf 12}, 81 (1956) ; R.-S. Cahn, {\it J. Chem. Educ.\/},  {\bf 41}, 116 (1964) ; {\it Erratum, ibid.\/}, 508 ; R.-S. Cahn, C.-K. Ingold and V. Prelog, {\it Ang. Chem.}, {\bf 5}, 385 (1966)
\bibitem{Craig:76} D.P. Craig and D.P. Mellor, {\it Discriminating Interactions Between Chiral Molecules\/} in {\it Topic in current Chemistry\/}, {\bf 63}, 1 (Springer-Verlag, Berlin, 1976).
\bibitem{Kornyshev:07} A.A. Kornyshev, D.J. Lee, S. Leikin and A. Wynveen {\it Rev. Mod. Phys.\/}, {\bf 79}, 943 (2007).
\bibitem{Kuhn:08} H. Kuhn, {\it Curr. Op. Coll. \& Int. Sci.\/}, {\bf 13}, 3 (2008).
\bibitem{Meyer:75}R.B. Meyer, L.Li\'ebert, L. Strzelecki and P. Keller, {\it J. de Physique}, {\bf 36}, L69 (1975).
\bibitem{vanderMeer:76} B.W. van der Meer, G. Vertogen, A.J. Dekker and J.G.J. Ypma, {\it J. Chem. Phys.\/}, {\bf 65}, 3935 (1976). 
\bibitem{Goossens:89} W.J.A. Goossens, {\it Phys. Rev. A\/}, {\bf 40}, 4019 (1989).
\bibitem{Chandrasekhar:92} S. Chandrasekhar, {\it Liquid Crystals, 2nd ed.\/}, (Cambridge Universty Press, Cambridge, 1992).
\bibitem{deGennes:93} P.G. de Gennes and J. Prost, {\it The Physics of Liquid Crystals, 2nd ed.\/}, (Clarendon Press, Oxford, 1993). 
\bibitem{Kamien:01} R.D. Kamien and J.V. Selinger, {\it J. Phys.: Condens. Matter\/}, {\bf 13}, R1 (2001). 
\bibitem{Emelyanenko:06} A.V. Emelyanenko, A. Fukuda and J.K. Vij, {\it Phys. Rev. E\/}, {\bf 74}, 011705 (2006).
\bibitem{Xu:01} J. Xu, R.L.B. Selinger, J.V. Selinger and R. Shashidhar, {\it J. Chem. Phys.\/}, {\bf 115}, 4333 (2001).
\bibitem{Yan:08} F. Yan, C. Hixson and D. Earl, {\it Phys. Rev. Lett.\/}, {\bf 101}, 157801 (2008).
\bibitem{deJeu:03} W.H. de Jeu, B.I. Ostrovskii and A.N. Shalaginov, {\it Rev. Mod. Phys.\/}, {\bf 75}, 181 (2003).
\bibitem{Osipov:96} M.A. Osipov, H. Stegemeyer and A. Sprick, {\it Phys. Rev. E\/}, {\bf 54}, 6387 (1996). 
\bibitem{Cepic:06} M. \u{C}epi\u{c}, {\it Europhys. Lett.}, {\bf 75}, 771 (2006).
\bibitem{Kuhn:30} W. Kuhn, {\it Trans. Farraday Soc.\/}, {\bf 26}, 293 (1930).
\bibitem{Caldwell:71} D.J. Caldwell and H. Eyring, {\it The Theory of Optical Activity.\/}, (New-York, Wiley Interscience, 1971).
\bibitem{Osipov:95} M.A. Osipov, B.T. Pickup and D.A. Dunmur, {\it Mol. Phys.\/}, {\bf 84}, 1193 (1995).
\bibitem{Wang:05} X.-O. Wang, J.-Q. Li and C.-F. Li, {\it Chem. Phys.\/}, {\bf 320}, 37 (2005).
\bibitem{Wang:08} X.-O. Wang, L.-J. Gong and C.-F. Lei, {\it J. Chem. Phys.\/}, {\bf 129}, 074708 (2008).
\bibitem{Maki:96} J.J. Maki and A. Persoons, {\it J. Chem. Phys.\/}, {\bf 104}, 9340 (1996). 
\bibitem{Trost:07} J.Trost and K. Hornberger, {\it Chem. Phys.\/}, {\bf 335}, 115 (2007).
\bibitem{Barros:06} E.B. Barros {\it et al\/}, {\it Phys. Rep.\/}, {\bf 431}, 261 (2006). 
\bibitem{Anantram:06} M.P. Anantram and F.L\'eonard {\it Rep. Prog. Phys.\/}, {\bf 69}, 507 (2006).
\bibitem{Charlier:07} J. Charlier, X. Blase  and S. Roche, {\it Rev. Mod. Phys.\/}, {\bf 79}, 677 (2007).
\bibitem{Lee:04} D.J. Lee, A. Wynveen and A.A. Kornyshev, {\it Phys. Rev. E\/}, {\bf 70}, 051913 (2004).  
\bibitem{Grason:07} G.M. Grason and R.F. Bruinsma, {\it Phys. Rev. E\/}, {\bf 76}, 021924 (2007).
\bibitem{note1} D.P. Craig and D. P. Mellor \cite{Craig:76} have chosen the term {\it chirodiastaltic} ('diastaltic' $\equiv$ 'serving to distinguish') because of its explicit reference to chirality. Another terminology {\it diastereotopic} has been introduced by B. Bosnich and D.W. Watts B. Bosnich and D.W. Watts, {\it Inorg. Chem.\/}, {\bf 14}, 47 (1975).
\bibitem{Barron:87} L. D. Barron and C. J. Johnston,  {\it Mol. Phys.\/}, {\bf 42}, 33 (1981).
\bibitem{Jenkins:94} J. K. Jenkins, A. Salam and T. Thirunamachandran,  {\it Mol. Phys.\/}, {\bf 82}, 835 (1994).
\bibitem{Yang:87} D.K. Yang and P.P. Crooker, {\it Phys. Rev. A\/}, {\bf 35}, 4419 (1987).  
\bibitem{Evans:92} G.T. Evans, {\it Mol. Phys.\/}, {\bf 77}, 969 (1992).
\bibitem{Ferrarini:96} A. Ferrarini, G.J. Moro and P.L. Nordio, {\it Mol. Phys.\/}, {\bf 87}, 485 (1996).
\bibitem{Berardi:03} R. Berardi, M. Cecchini and C. Zannoni, {\it J. Chem. Phys.\/}, {\bf 119}, 9933 (2003).
\bibitem{Cao:05} M. Cao and P.A. Monson, {\it J. Chem. Phys.\/}, {\bf 122}, 054505 (2005).
\bibitem{Peon:06} J. Pe\'on, J. Saucedo-Zugazagoitia, F. Pucheta-Mendez, R.A. Perusquia, G. Sutmann and J. Quintana-H, {\it J. Chem. Phys.\/}, {\bf 125}, 104908 (2006).
\bibitem{Paci:01} I. Paci and N.M. Cann, {\it J. Chem. Phys.\/}, {\bf 115}, 8489 (2001).
\bibitem{Paci:03} I. Paci, J. Dunford and N.M. Cann, {\it J. Chem. Phys.\/}, {\bf 118}, 7519 (2003).
\bibitem{Paci:04} I. Paci and N.M. Cann, {\it J. Chem. Phys.\/}, {\bf 120}, 4816 (2004).
\bibitem{Huh:04} Y. Huh and N.M. Cann, {\it J. Chem. Phys.\/}, {\bf 121}, 10299 (2004).
\bibitem{Memmer:01} R. Memmer, {\it J. Chem. Phys.\/}, {\bf 114}, 8210 (2001).
\bibitem{Craig:74} D.P. Craig and P.E. Schipper, {\it Chem. Phys. Lett.\/}, {\bf 25}, 476 (1974).
\bibitem{Craig:75} D.P. Craig and P.E. Schipper, {\it Proc. R. Soc. London, Ser. A\/}, {\bf 342}, 19 (1975).
\bibitem{note2} An example of an achiral model of dipolar hard spheres possessing two colinear dipoles is given in the reference : P.J. Camp and G.N. Patey, {\it Phys. Rev. E\/}, {\bf 60}, 4280 (1999).
\bibitem{Luzanov:01} A.V. Luzanov and L.N. Lisetskii {\it J. Struct. Chem.\/}, {\bf 42}, 544 (2001).
\bibitem{Bosello:03} C.A. Bosello and R. Nibbi, {\it Math. Meth. Appl. Sci.\/}, {\bf 26}, 375 (2003).
\bibitem{Salam:06} A. Salam, {\it J. Chem. Phys.\/}, {\bf 124}, 014302 (2006).
\bibitem{Choi:07} J. Choi and M.Cho, {\it J. Chem. Phys.\/}, {\bf 127}, 024507 (2007).
\bibitem{Weis:05c} J.-J.  Weis and D. Levesque, {\it Advanced Computer Simulation Approaches for Soft Matter Sciences II\/}, edited by C. Holm and K. Kremer, Advances in Polymer Science Vol. 185 (Springer, New York, 2005).
\bibitem{Jackson:75} J.D. Jackson, {\it Classical Electrodynamics, 2nd ed.\/} (Wiley, New York, 1975)
\bibitem{Gray:84} C.G. Gray and K.E. Gubbins, {\it Theory of molecular fluids. Volume 1: Fundamentals.\/} (Clarendon Press, Oxford, 1984).
\bibitem{Harris:99} A.B. Harris, R.D. Kamien and T.C. Lubensky, {\it Rev. Mod. Phys.\/}, {\bf 71}, 1745 (1999).
\bibitem{DeLeeuw:80} De Leeuw, S.W., Perram, J.W., and Smith, E.R., 1980, {\it Proc. R. Soc. Lond. A\/}, {\bf 373}, 27 ; $ibid$, 57.
\bibitem{Weis:93a} J.-J. Weis and D. Levesque, {\it Phys. Rev. Lett.\/}, {\bf 71}, 2729 (1993).
\bibitem{Weis:93b} J.-J. Weis and D. Levesque, {\it Phys. Rev. E\/}, {\bf 48}, 3728 (1993).
\bibitem{Levesque:94} D. Levesque and J.-J. Weis, {\it Phys. Rev. E\/}, {\bf 49}, 5131 (1994).
\bibitem{Lomba:00} E. Lomba, F. Lado and J.-J. Weis, {\it Phys. Rev. E\/}, {\bf 61}, 3838 (2000).
\bibitem{Weis:02} J.-J. Weis, {\it Mol. Phys.\/}, {\bf 100}, 579 (2002).
\bibitem{Weis:02a} J.-J. Weis, J.M. Tavares and M.M. Telo da Gama, {\it J. Phys: Condens. Matter\/}, {\bf 14}, 9171 (2002).
\bibitem{Tavares:02} J.M. Tavares, J.-J. Weis and M.M. Telo da Gama, {\it Phys. Rev. E\/}, {\bf 65}, 061201 (2002).
\bibitem{Weis:03} J.-J. Weis, {\it J. Phys: Condens. Matter\/}, {\bf 15}, S1471 (2003).
\bibitem{Fernaud:03a} M.J. Fernaud, E. Lomba, J.-J. Weis and D. Levesque, {\it Mol. Phys.\/}, {\bf 101}, 1721 (2003).
\bibitem{Fernaud:03b} M.J. Fernaud, E. Lomba, C. Martin, D. Levesque and J.-J. Weis {\it J. Chem. Phys.\/}, {\bf 119}, 364 (2003).
\bibitem{Weis:05a} J.-J. Weis, {\it Mol. Phys.\/}, {\bf 103}, 7 (2005).
\bibitem{Weis:05b} J.-J. Weis, {\it J. Chem. Phys.\/}, {\bf 123}, 044503 (2005).
\bibitem{Holm:05} C. Holm and J.-J. Weis, {\it Curr. Op. Coll. \& Int. Sci.\/}, {\bf 10}, 133 (2005).
\bibitem{Tavares:06} J.M. Tavares, J.-J. Weis and M.M. Telo da Gama, {\it Phys. Rev. E\/}, {\bf 73}, 041507 (2006).
\bibitem{Weis:06} J.-J. Weis and D. Levesque, {\it J. Chem. Phys.\/}, {\bf 125}, 034504 (2006).
\bibitem{Alvarez:08} C. Alvarez, M. Mazars and J.-J. Weis, {\it Phys. Rev. E\/}, {\bf 77}, 051501 (2008).
\bibitem{Richardi:08} J. Richardi, M.P. Pileni and J.-J. Weis, {\it Phys. Rev. E\/}, {\bf 77}, 061510 (2008).
\bibitem{Wertheim:71} M.S. Wertheim, {\it J. Chem. Phys.\/}, {\bf 55}, 4291 (1971).
\bibitem{Blum:72} L. Blum and A.J. Torruella, {\it J. Chem. Phys.\/}, {\bf 56}, 303 (1972).
\bibitem{Frenkel:04} D. Frenkel, {\it Proc. Natl. Acad. Sci. U.S.A\/}, {\bf 101}, 51 (2004).
\bibitem{Whitelam:07} S. Whitelam and P.L. Geissler, {\it J. Chem. Phys.\/}, {\bf 127}, 154101 (2007).
\bibitem{Almarza:07} N.G. Almarza and E. Lomba, {\it J. Chem. Phys.\/}, {\bf 127}, 084116 (2007).
\bibitem{Nelson:81} D.R. Nelson and J. Toner, {\it Phys. Rev. B\/}, {\bf 24}, 363 (1981) ; P.J. Steinhardt, D.R. Nelson and M. Ronchetti, {\it Phys. Rev. B\/}, {\bf 28}, 784 (1983). 
\bibitem{Aguado:03} A. Aguado and P.A. Madden, {\it J. Chem. Phys.\/}, {\bf 119}, 7471 (2003).
\bibitem{Laino:08} T. Laino and J. Hutter, {\it J. Chem. Phys.\/}, {\bf 129},  074102 (2008).
\bibitem{Vieillard:72} J. Vieillard-Baron, {\it J. Chem. Phys.\/}, {\bf 56}, 4729 (1972).
\bibitem{Frenkel:85} D. Frenkel and B.M. Mulder, {\it Mol. Phys.\/}, {\bf 55}, 1171 (1985) ; {\it ibid.}, 1193 (1985).
\bibitem{Vieillard:74} J. Vieillard-Baron, {\it Mol. Phys.\/}, {\bf 28}, 809 (1974).
\bibitem{Veerman:90} J.A.C. Veerman and D. Frenkel, {\it Phys. Rev. A\/}, {\bf 41}, 3237 (1990) ; {\it ibid.}, {\bf 43}, 1193 (1985).
\bibitem{Gay:81} J.G. Gay and B.J. Berne, {\it J. Chem. Phys.\/}, {\bf 74}, 3316 (1981).
\bibitem{deMiguel:92} E. de Miguel, L.F. Rull and K.E. Gubbins, {\it Phys. Rev. A\/}, {\bf 44}, 3813 (1992)

\end{thebibliography}
\end{document}